\documentclass[longauth]{aa}

\usepackage{graphicx}
\usepackage{txfonts}
\usepackage{gensymb,amssymb,amsmath}
\usepackage{xcolor}
\usepackage[version=3]{mhchem}
\usepackage{caption,subcaption}
\usepackage{multirow}

\usepackage[colorlinks = true,
            linkcolor = blue,
            urlcolor  = blue,
            citecolor = blue,
            anchorcolor = blue]{hyperref}

\begin{document}

\title{MINDS. Water reservoirs of compact planet-forming dust disks}
\subtitle{A diversity of \ce{H_2O} distributions}
\author{Milou Temmink\orcid{0000-0002-7935-7445}\inst{1} \and
        Andrew D. Sellek\orcid{0000-0003-0330-1506}\inst{1} \and
        Danny Gasman\orcid{0000-0002-1257-7742}\inst{2} \and
        Ewine F. van Dishoeck\orcid{0000-0001-7591-1907}\inst{1,3} \and
        Marissa Vlasblom\orcid{0000-0002-3135-2477}\inst{1} \and
        Ang\`el Pranger\inst{1} \and
        Manuel G\"udel\orcid{0000-0001-9818-0588}\inst{4,5} \and
        Thomas Henning\orcid{0000-0002-1493-300X}\inst{6} \and
        Pierre-Olivier Lagage\inst{7} \and
        Alessio Caratti o Garatti\orcid{0000-0001-8876-6614}\inst{8,9} \and
        Inga Kamp\orcid{0000-0001-7455-5349}\inst{10} \and
        G\"oran Olofsson\orcid{0000-0003-3747-7120}\inst{11} \and 
        Aditya M. Arabhavi\orcid{0000-0001-8407-4020}\inst{10} \and
        Sierra L. Grant\orcid{0000-0002-4022-4899}\inst{12} \and
        Till Kaeufer\orcid{0000-0001-8240-978X}\inst{13} \and
        Nicolas T. Kurtovic\orcid{0000-0002-2358-4796}\inst{3} \and
        Giulia Perotti\orcid{0000-0002-8545-6175}\inst{14,6} \and
        Matthias Samland\orcid{0000-0001-9992-4067}\inst{6} \and
        Kamber Schwarz\orcid{0000-0002-6429-9457}\inst{6} \and
        Beno\^it Tabone\orcid{0000-0002-1103-3225}\inst{15} 
        }
\institute{Leiden Observatory, Leiden University, PO Box 9513, 2300 RA Leiden, the Netherlands \\
          \email{temmink@strw.leidenuniv.nl} \and
          Institute of Astronomy, KU Leuven, Celestijnenlaan 200D, 3001 Leuven, Belgium \and
          Max-Planck-Institut f\"ur Extraterrestrische Physik, Giessenbachstraße 1, D-85748 Garching, Germany \and
          Dept. of Astrophysics, University of Vienna, T\"urkenschanzstr. 17, A-1180 Vienna, Austria \and
          ETH Z\"urich, Institute for Particle Physics and Astrophysics, Wolfgang-Pauli-Str. 27, 8093 Z\"urich, Switzerland \and
          Max-Planck-Institut f\"{u}r Astronomie (MPIA), K\"{o}nigstuhl 17, 69117 Heidelberg, Germany \and
          Universit\'e Paris-Saclay, Universit\'e Paris Cit\'e, CEA, CNRS, AIM, F-91191 Gif-sur-Yvette, France \and
          INAF – Osservatorio Astronomico di Capodimonte, Salita Moiariello 16, 80131 Napoli, Italy \and
          Dublin Institute for Advanced Studies, 31 Fitzwilliam Place, D02 XF86 Dublin, Ireland \and
          Kapteyn Astronomical Institute, Rijksuniversiteit Groningen, Postbus 800, 9700AV Groningen, The Netherlands \and
          Department of Astronomy, Stockholm University, AlbaNova University Center, 10691 Stockholm, Sweden \and
          Earth and Planets Laboratory, Carnegie Institution for Science, 5241 Broad Branch Road, NW, Washington, DC 20015, USA \and
          Department of Physics and Astronomy, University of Exeter, Exeter EX4 4QL, UK \and
          Niels Bohr Institute, University of Copenhagen, NBB BA2, Jagtvej 155A, 2200 Copenhagen, Denmark \and
          Universit\'e Paris-Saclay, CNRS, Institut d’Astrophysique Spatiale, 91405, Orsay, France}
\date{Received 21/02/2025; accepted 20/05/2025}


\abstract
{Millimetre-compact dust disks are thought to have efficient radial drift of icy dust pebbles, which has been hypothesised to produce an enhanced cold ($T<$400 K) \ce{H_2O} reservoir in their inner disks. Mid-infrared spectral surveys, now with the \textit{James Webb} Space Telescope (JWST), pave the way to explore this hypothesis. In this work, we test this theory for 8 compact disks ($R_\mathrm{dust}<$60 au) with JWST-MIRI/MRS observations.}
{To explore the \ce{H_2O} distribution in the inner disk and whether these disks are enhanced in cold \ce{H_2O} emission, we analyse the different reservoirs that can be probed with the pure rotational lines ($>$10 $\mathrm{\mu}$m) by JWST: hot ($T>800$ K), intermediate (400$<T<$800 K), and cold ($T<$400 K). }
{We probe the \ce{H_2O} reservoirs with JWST-MIRI observations for a sample of 8 compact disks through parametric column density profiles (power laws, jump abundances, and parabolas), multiple component (two or three) slab models, and line flux ratios.}
{We find that not all compact disks show strong enhancements of the cold \ce{H_2O} reservoir, instead we propose three different classes of inner disk \ce{H_2O} distributions. Four of our disks (BP~Tau, CY~Tau, DR~Tau, and RNO~90; Type N or ``Normal'' disks) appear to have similar \ce{H_2O} distributions as many of the large and structured disks, as is indicated by the slab model fitting and the line flux ratios. These disks have a small cold reservoir, suggesting the inward drift of dust, but it is not as efficient as hypothesised before. Only two disks (FT~Tau and XX~Cha; Type E or cold \ce{H_2O} enhanced disks) do show a strong enhancement of the cold \ce{H_2O} emission, agreeing with the original hypothesis. The two remaining disks (CX~Tau and DN~Tau; Type P or \ce{H_2O}-poor disks) are found to be very \ce{H_2O}-poor, yet show emission from either the hot or immediate reservoirs (depending on the fit) in addition to emission from the cold one.  For the three types, we find that different parametrisations are able to provide a good description of the observed \ce{H_2O} spectra; a jump abundance at a free temperature is amongst the preferred profiles for all three types, suggesting that this profile can provide a good description of the observed reservoirs for most disks. The multiple component analysis yields similar results as the parametric models. However, in some cases, a power law can give an entirely different distribution compared to the other parametric models. Finally, we also report the detection of other molecules in these disks, including a tentative detection of \ce{CH_4} in CY~Tau.}
{Not all compact disks follow the hypothesis that their cold \ce{H_2O} reservoir is enhanced following efficient radial drift. Therefore, we introduced a classification based on the observed \ce{H_2O} reservoirs, which should hold for all (isolated) disks: Type N, Type E, and Type P. Type N disks are considered to behave as many other (large and structured) disks, with all three reservoirs present, yet the cold emission is not enhanced. The Type E disks show strong enhancements of the cold \ce{H_2O} emission, while the Type P disks are generally \ce{H_2O}-poor.}

\keywords{astrochemistry - protoplanetary disks - stars: variables: T-Tauri, Herbig Ae/Be - infrared: general}


\maketitle

\section{Introduction} \label{sec:Intro}
\begin{table*}[ht!]
    \centering
    \caption{Stellar properties and observational details of the sample of compact disks studied in this work.}
    \begin{tabular}{c c c c c c c c c c c}
        \hline\hline
        Source & $M_*$ & $L_*$ & Dist. & $v_\textnormal{hel}$ & Inclination & $R_\mathrm{in}$ & $R_\mathrm{dust,95\%}$ & Obs. date & Int. time & References \\
        & [M$_\odot$] & [L$_\odot$] & [pc] & [km~s$^{-1}$] & [\degree] & [au] & [au] & dd/mm/yyyy & [min] & $M_*$/$L_*$/$v_\textnormal{hel}/i/R_\mathrm{dust}$ \\
        \hline
        BP~Tau & 0.52 & 0.40 & 129 & 16.76 & 23.6 & 0.04 & 41 & 20/02/2023 & 29.0 & $m1,l1,v1,i1,d1$ \\
        CX~Tau & 0.37 & 0.26 & 128 & 19.30 & 55.1 & 0.04 & 38 & 20/02/2023 & 60.0 & $m2,l2,v2,i2,d2$ \\
        CY~Tau & 0.31 & 0.26 & 128 & 15.10 & 32.0 & 0.04 & $<$64 & 20/02/2023 & 60.0 & $m2,l2,v2,i3,d3$ \\
        DN~Tau & 0.52 & 0.70 & 128 & 18.80 & 35.2 & 0.06 & 61 & 28/09/2023 & 60.0 & $m1,l1,v2,i4,d1$ \\
        DR~Tau & 0.93 & 0.63 & 195 & 27.60 & 5.4 & 0.06 & 54 & 04/03/2023 & 27.8 & $m1,l1,v3,i4,d1$ \\
        FT~Tau & 0.34 & 0.15 & 127 & 17.24 & 35.5 & 0.03 & 45 & 20/02/2023 & 30.0 & $m1,l1,v4,i4,d1$ \\
        RNO~90 & 1.50 & 2.14 & 117 & -10.10 & 37.0 & 0.10 & $<$47 & 25/08/2023 & 33.4 & $m3,l3,v5,i5,d3$ \\
        XX~Cha & 0.36 & 0.29 & 190 & 16.26 & 39.7 & 0.04 & $<$25 & 31/08/2023 & 60.0 & $m4,l4,v6,i6,d3$ \\
        \hline
    \end{tabular}
    \label{tab:Sample}
    \tablefoot{All distances have been acquired from GAIA \citep{GC18}. References (mass, luminosity, heliocentric velocity, inclination, dust radii): $m1$-\citet{LongEA19}, $m2$-\citet{SimonEA17}, $m3$-\citet{BanzattiEA22}, $m4$-\citet{ManaraEA16}, $l1$-\citet{LongEA19}, $l2$-\citet{HH14}, $l3$-\citet{BanzattiEA22}, $l4$-\citet{ManaraEA16}, $v1$-\citet{FangEA18}, $v2$-\citet{BanzattiEA19}, $v3$-\citet{ArdilaEA02}, $v4$-\citet{RoccatagliataEA20}, $v5$-\citet{BanzattiEA22}, $v6$-\citet{NguyenEA12}, $i1$-\citet{GasmanEA25}, $i2$-\citet{FacchiniEA17}, $i3$-\citet{SimonEA17}, $i4$-\citet{LongEA19}, $i5$-\citet{BosmanEA21}, $i6$-This work, $d1$-\citet{LongEA19}, $d2$-\citet{FacchiniEA19}, $d3$-Estimated in this work}
\end{table*}
The inner regions ($<$10 au) of planet-forming disks are active sites of planet formation \citep{MorbidelliEA12,DJ18}. The chemical composition of these forming planets is set by the elemental and molecular composition of the nascent disk. One of the key ingredients for habitable worlds is \ce{H_2O} and, therefore, studying the available \ce{H_2O} vapour reservoirs in disks is of great importance. \\
\indent Based on \ce{H_2O} vapour observations with the \textit{Spitzer} Space Telescope and spatially resolved continuum images with the Atacama Large Millimeter/submillimeter Array, \citet{BanzattiEA20} proposed scenarios for the expected \ce{H_2O} reservoirs given the sizes of the dust disks. Disks that are observed to be compact in their millimetre emission ($R_\mathrm{dust}<$60 au) are thought to have efficient radial drift, leading to enriched \ce{H_2O} reservoirs ($N\sim$10$^{18}$-10$^{21}$ cm$^{-2}$) in the innermost regions ($<$ few au) of these disks. The larger, structured disks (60$<R_\mathrm{dust}<$300 au) are expected to not show these enriched \ce{H_2O} abundances ($\leq$10$^{17.5}$ cm$^{-2}$), since substructures (i.e., gaps and rings) may trap the icy dust grains in the outer regions, halting them from reaching the inner disk and from crossing the \ce{H_2O} snowline. Furthermore, for disks with large cavities, the situation is expected to be even more extreme, with the inner disks expected to be depleted of \ce{H_2O}. \\
\indent Recent modelling works have explored the notion of  both unimpeded enhancement due to radial drift and the opposite case of substructures halting the inward drift of pebbles. In particular, \citet{KalyaanEA23} investigated the influence of gaps on the gaseous enhancement in the inner disk. They note that the enhancement, at least for the \ce{H_2O} vapour reservoir, is of limited duration (up to a few million years), before the material gets accreted onto the host star. Only when the gaps do not block the dust entirely, but some of the grains are still able to pass through \citep{PinillaEA24,MahEA24}, the lifetime of this enhanced reservoir can be prolonged. However, the inward drift of the dust particles does not only enhance the gaseous reservoir, it also increases the opacity of the dust itself. Recent models by \citet{SellekEA24} suggest that the increased opacity of the dust elevates the $\tau$=1 layer of the continuum and, therefore, may hide a larger gas reservoir that lies deeper in the disk. The observable column densities may thus not reflect the enhanced reservoirs.  Further work by \citet{HougeEA25} supports this idea and also investigated the competition between photodissociation and vertical mixing on the available reservoir. Finally, \citet{KalyaanEA23}, \citet{SellekEA24}, and \citet{EasterwoodEA24} note the importance of the gap locations and the time at which they form for the drifting pebbles. Gaps located at smaller radial distances and which formed early on are more effective at limiting the gaseous enhancement, as more dust grains are blocked outside the snowline. However, they may leak more for a given gap depth or grain size. For gaps located at larger radial distances, the enhancement is stronger, as a smaller icy dust reservoir is blocked. Furthermore, the way the gap opens may also be of importance. \citet{LienertEA24} have shown that gaps opened due to internal photoevaporation are able to block both the pebble and gas flow, strongly influencing the inner disk composition. In contrast, models by \citet{GreenwoodEA19} have shown that, in the absence of gaps, the dust opacity decreases as the disk evolves with time and this leads to an increase in the \ce{H_2O} flux. Current observations of mid-infrared spectra of large and structured disks with the \textit{James Webb} Space Telescope (JWST), including those with large cavities, already show that inner regions of such disks are not completely void of \ce{H_2O} \citep{PerottiEA23,SchwarzEA24} and, in some cases, they may have strong emission from the cold reservoir ($<$400 K; \citealt{GasmanEA23,GasmanEA25}). \\
\indent The \textit{James Webb} Space Telescope (\citealt{RigbyEA23}) provides the best opportunity to fully explore the available \ce{H_2O} reservoirs in the inner regions of planet-forming disks, using the increased sensitivity and resolution of the Medium Resolution Spectrometer (MRS; \citealt{WellsEA15,ArgyriouEA23}) of the Mid-InfraRed Instrument (MIRI; \citealt{WrightEA15,RiekeEA15,WrightEA23}) with respect to the \textit{Spitzer} Space Telescope. Since its launch, multiple works have studied the available reservoirs, including the analysis of \ce{H_2O} emission across the entire JWST-MIRI wavelength range \citep{GasmanEA23} using 0D local thermal equilibrium (LTE) slab models. This analysis directly proved the existence of an expected radial temperature gradient, where the longer wavelengths probe larger radii and, thus, colder temperatures \citep{BlevinsEA16,BanzattiEA17,BanzattiEA23a,BanzattiEA24}. \citet{BanzattiEA23b} identified the emergence of a cold \ce{H_2O} reservoir ($T<$240 K), which is expected to be the effect of radial dust drift. Subsequently, multiple slab models were fitted to the pure rotational \ce{H_2O} spectra ($>10$ $\mathrm{\mu}$m) identifying the available reservoirs \citep{PontoppidanEA24,TemminkEA24b}, and a third cold component, in addition to hot ($>$800 K) and warm ($400<T<800$ K components, was found to be necessary to describe the same cold \ce{H_2O} reservoir of $T<$240 K. The three-component analysis of \citet{TemminkEA24b} also provided another confirmation of the radial temperature gradient, where the multiple component analysis of DR~Tau could be approximated with a temperature profile of $T(R)\sim500\left(R/1\mathrm{au}\right)^{-0.5}$ K. More recently, efforts have been made to increase the complexity by moving away from using multiple components and to describe the temperature and column density profiles by parametric functions \citep{KeauferEA24,RomeroMirzaEA24}. These profiles, including simple and exponentially tapered power laws, show that the pure rotational \ce{H_2O} transitions can be explained by such parametric models, but this has only been tested for a limited sample of disks with different disk characteristics, such as their radial sizes and structures. We also note the effort by \citet{WoitkeEA19}, who, for the first time, used a full 2D thermochemical code to model the observed molecular emission in the outbursting source of EX~Lup. \\
\indent In this work, we aim to analyse the pure rotational \ce{H_2O} emission in 8 millimetre-compact disks (dust radii of $R_\mathrm{dust,95\%}\sim25-60$ au; \citealt{LongEA19,FacchiniEA19}), using both the parametric and multiple component techniques highlighted above. Our criterion for a dust disk to be compact follows the classification of \citet{BanzattiEA20}, i.e., compact disks have dust radii $\lesssim$60 au. All sources have been observed with JWST/MIRI-MRS as part of the JWST Guaranteed Time Observations Program MIRI mid-INfrared Disk Survey (MINDS, PID:1282, PI: T. Henning; \citealt{KampEA23,HenningEA24}). Throughout our analysis, we aim to investigate the strengths of the different \ce{H_2O} reservoirs, indicated throughout the paper as hot ($>$800 K), intermediate (400-800 K), and cold ($<$400 K, see also \citealt{BanzattiEA23b}). We use this to explore the \ce{H_2O} reservoirs and column density profiles. In addition, we also infer whether these compact disks show an enhancement in the cold \ce{H_2O} reservoir, as suggested by \citet{BanzattiEA20}, or whether the situation is more complex than this scenario. \\
\indent The paper is structured as follows: Section \ref{sec:Obs} describes the sample and the observations, while Section \ref{sec:Anal} contains a description of the used methodology. The results are represented in Section \ref{sec:Res}, and are interpreted in Section \ref{sec:Disc}. Aside from the interpretation, Section \ref{sec:Disc} also discusses the role of substructures in setting the \ce{H_2O} reservoirs and makes a comparison with existing models. Finally, Section \ref{sec:CS} contains our conclusions and a short summary. 

\section{Sample and observations} \label{sec:Obs}
\begin{figure*}[ht!]
    \centering
    \includegraphics[width=\textwidth]{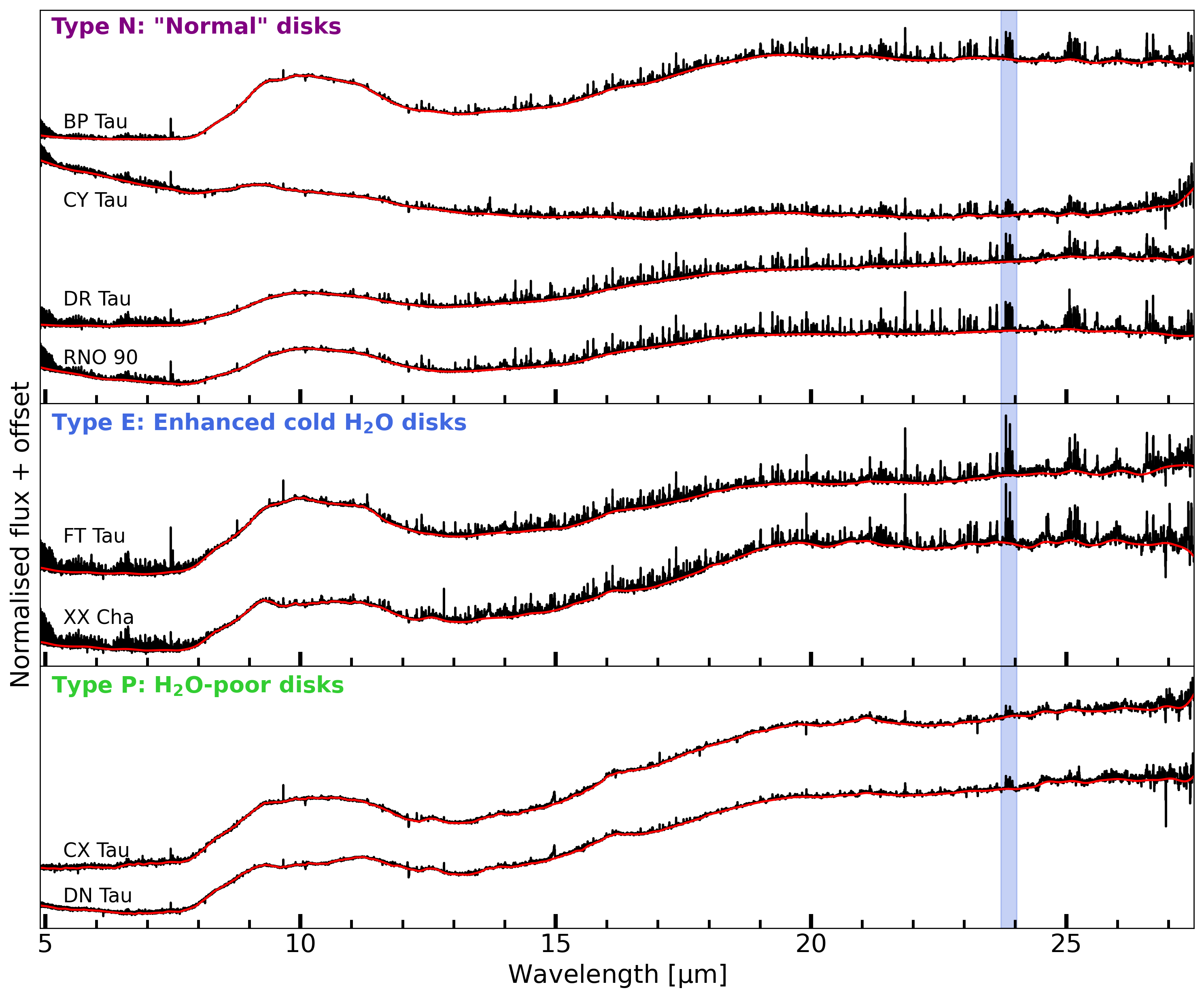}
    \caption{Normalised to peak flux spectra of our sample of millimetre-compact disks. The red line indicates the estimated continuum. The indicated types and their meaning are discussed in Section \ref{sec:Types}. In blue we have highlighted the 23.72-24.03 $\mathrm{\mu}$m wavelength region, where two transitions are located that are most important for obtaining information about the cold \ce{H_2O} reservoir.}
    \label{fig:Sample}
\end{figure*}
\subsection{Sample}
Our sample of compact planet-forming disks consists of eight sources: BP~Tau, CX~Tau, CY~Tau, DN~Tau, DR~Tau, FT~Tau, RNO~90, and XX~Cha, for which stellar properties are summarised in Table \ref{tab:Sample}. Dedicated papers, exploring the observed molecular emission, exist for two of the sources: CX~Tau \citep{VlasblomEA24b} and DR~Tau \citep{TemminkEA24,TemminkEA24b}. \\
\indent Although the disks in our sample are known to be rather compact in the millimetre continuum emission ($R_\mathrm{dust}\lesssim$60 au), limited information is known about their gas disk sizes ($R_\mathrm{gas}$). \citet{TrapmanEA19} proposed that a gas-to-dust size ratio of $R_\mathrm{gas}/R_\mathrm{dust}>$4 is a clear sign of radial drift. For our sources, only one disk has literature values for both $R_\mathrm{dust}$ and $R_\mathrm{gas}$: CX~Tau \citep{FacchiniEA19}. These values suggest a ratio of $R_\mathrm{gas}/R_\mathrm{dust}>$5, hinting at strong radial drift being the potential cause of this disk's compactness. For the other disks, the gas radius has not been determined and, in some cases, high-resolutions observations of the continuum emission do not (yet) exist. Consequently, the gas-to-dust size ratio is still to be determined in a consistent manner and it is unclear whether the compactness of these disks can be attributed to an evolution dominated by radial drift or simply to their initial conditions. For these reasons, we have chosen our sample to follow the criteria of \citet{BanzattiEA20} and consider the disks with millimetre continuum emission radii of $R_\mathrm{dust}\lesssim$60 au to be compact. \\
\indent As can be seen from Table \ref{tab:Sample}, our sample is rather homogeneous in terms of masses, luminosities, and inclinations. There are a few outliers: i.e., DR~Tau and RNO~90 have higher stellar masses compared to the others, while DR~Tau also has the lowest inclination ($i\sim$5.4\degree, \citealt{LongEA19}). CX~Tau, on the other hand, has the highest inclination ($\sim$55.1\degree). Furthermore, as 6 out of 8 disks are located in the Taurus star-forming region, we can expect them to be of similar ages ($\sim$1-3 Myr, \citealt{KroliEA21,Luhman23}). In contrast, RNO~90 is located in Ophiuchus and may, therefore, be on the younger side. These small differences suggest that our sample is overall rather homogeneous and differences must therefore be due to either their initial conditions or differences in their (potentially drift-dominated) evolution. A larger sample will be analysed in Temmink et al. (in prep.), which may provide more insights into the differences related to the stellar properties.

\subsection{Observations and data reduction}
\indent The JWST-MIRI/MRS details (date of observation and integration time) are also included in Table \ref{tab:Sample}. All MRS spectra have been taken in the FASTR1 readout mode with a four-point dither pattern using all three grating settings (A, B, and C). All data have been reduced using a standard pipeline reduction (version 1.16.1; \citealt{BushouseEA24}) and pmap 1315. The spectra have been extracted through aperture photometry, where the aperture has a size of 2$\times$ the full width at half maximum (FWHM). Additionally, we have corrected for residual fringes using the implementation of the default pipeline, as most of the observations (except XX~Cha) were taken without target acquisition. The resulting spectra were continuum subtracted using an updated version of the method by \citet{TemminkEA24}, which makes use of the 'Iterative Reweighted Spline Quantile Regression' method included in the \textsc{pybaselines} package \citep{PyBaselines}. Before estimating the continuum, downward spikes were masked using the same method as in \citet{TemminkEA24}, but now they were masked per MIRI subband as opposed to over the whole spectrum. We used a quantile regression value of 0.1 and placed the knots of the cubic splines every 75 data points, except for Channel 4 and the silicate feature ($\sim$8.25-11.25 $\mathrm{\mu}$m), where the knots were placed every 25 points to better estimate the varying continuum. For CX~Tau and DN~Tau, a knot spacing of 25 data points was used for the entire spectrum, due to their strongly varying continuum. Given the large uncertainty of the observations at the location of the silicate feature, the quantile regression value was set to 0.5 for all disks, which places the baseline through the median of the observations. Additionally, we changed the continuum estimate for the $Q$-branches of \ce{CO_2}, \ce{HCN}, and \ce{C_2H_2}. The typical fit could overestimate the continuum of these $Q$-branches, effectively taking away molecular flux in the subtraction. To avoid this oversubtraction, we used a cubic interpolation of the estimated baseline just before and after the $Q$-branch to better fit the continuum level underneath the $Q$-branch. In particular, we masked the 13.45-14.20 $\mathrm{\mu}$m wavelength region that captures the \ce{HCN} and \ce{C_2H_2} $Q$-branches, and the 14.88-15.01 $\mathrm{\mu}$m region for that of \ce{CO_2}. The reduced spectra and continuum estimates are displayed in Figure \ref{fig:Sample}. \\
\indent Figure \ref{fig:Sample} shows that the shape of the spectra overall look similar and in some cases nearly indistinguishable. All sources have a silicate feature at $\sim$10 $\mathrm{\mu}$m and contain emission features from a variety of molecular species. The only outlier is CY~Tau, which has stronger emission at the shortest wavelengths. This is very likely due to an inner (puffy) rim self-shadowing the outer disk and, subsequently, lowering the mid- and far-infrared flux (see, for example, \citealt{DD04, WoitkeEA19b}). We leave the analysis of the dust continua for future work.

\section{Methodology} \label{sec:Anal}
To analyse the pure rotational \ce{H_2O} emission in the JWST-MIRI/MRS spectra of our sample of compact disks, we use slab models under the assumption of local thermal equilibrium (see \citealt{GrantEA23,TaboneEA23} for more details on the slab model generation). Following \citet{BanzattiEA24}, we include mutual line shielding in the \ce{H_2O} slab models to account for the shielding of ortho- and para-line pairs. Additionally, we take the mean spectral resolution, calculated using the results of \citet{PontoppidanEA24} and the updates by \citet{BanzattiEA24}, of each MIRI subband to properly sample the wavelength grid of the slabs over the entire spectrum. We use two different line widths for our models: a constant value of $\Delta V$=4.71 km s$^{-1}$, which is the line width of \ce{H_2} at a temperature of 700 K \citep{SalykEA11}, as well as a similar approach to \citet{RomeroMirzaEA24}, where the line width is taken as the sum in quadrature (hereafter simply called the quadrature line width) of the turbulent line width ($\Delta V_\mathrm{turb}$ fixed to 1.0 km s$^{-1}$) and the thermal line width ($\Delta V_\mathrm{therm}=\sqrt{2kT/m}$, with $T$ the temperature of the slab model and $m$ the mass of an \ce{H_2O} molecule). The two line widths are used to make a comparison between the practices used in the recent literature. While the earlier works (e.g., \citealt{TaboneEA23,GrantEA23,GasmanEA23}) have used a constant value of 4.71 km~s$^{-1}$, \citet{RomeroMirzaEA24} recently used the sum in quadrature. To investigate which one of these approaches fits better to our spectra, we test both options in our fits and make a comparison in Section \ref{sec:LWComp}. The python-package \textsc{spectres} \citep{SpectRes} is subsequently used to resample the slab models onto the wavelength grid of MIRI. However, before resampling, we apply a wavelength shift to the slab models (see also \citealt{PontoppidanEA24}), based on the heliocentric velocity of the sources (see Table \ref{tab:Sample}). Figure \ref{fig:VHelShift}, using the spectrum of DR~Tau as an example, shows that these wavelength shifts, although subtle, are required to properly recreate the observed line profiles and, subsequently, will improve fit results. \\
\indent We fit our spectra using two different approaches: similar to \citet{TemminkEA24b}, we use a multi-component analysis where three or two distinct components (with decreasing temperatures and increasing emitting areas) are fitted to the entire rotational spectrum. Furthermore, we use and extend upon the methods implemented by \citet{RomeroMirzaEA24}, who used (exponentially-tapered) power law profiles for the temperature and column density. We denote these models as multi-component and parametric models, where in the latter case we explore more options than (exponentially tapered) power laws. By using both these methods, we can explore the strengths of the different \ce{H_2O} reservoirs and investigate the robustness of what kind of profile for the column density is preferred in the inner region. By using both techniques, we also gain more insights into the changing excitation conditions with radius in the inner regions of these disks.

\subsection{Rotational \ce{H_2O} spectrum: parametric analysis} \label{sec:AnalH2O}
\begin{figure*}[ht!]
    \centering
    \includegraphics[width=\textwidth]{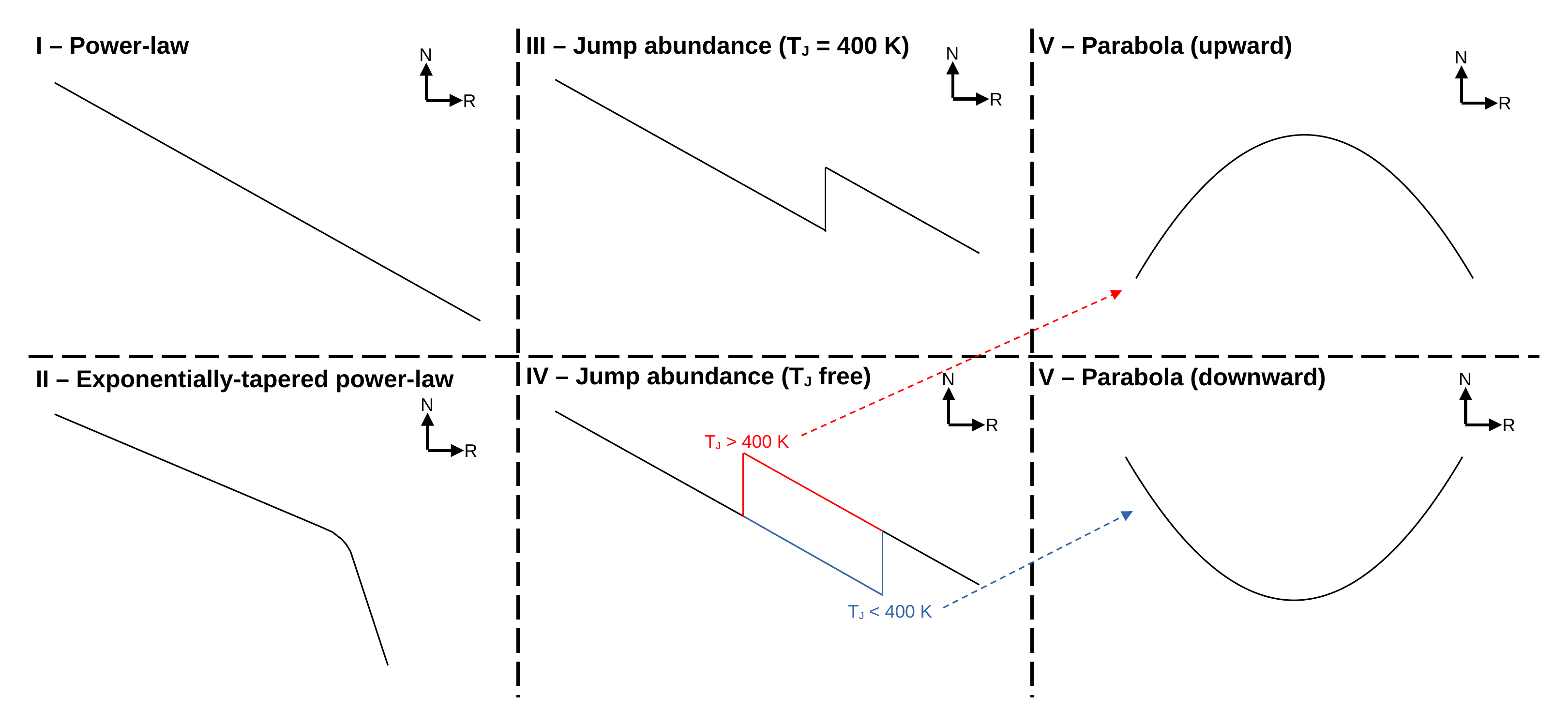}
    \caption{Schematic highlighting the different parametric models used for the column density profiles. Both the horizontal (radius) and vertical (column density) axes are taken in $\log_{10}$-space.}
    \label{fig:Schematic}
\end{figure*}
In an attempt to parametrically describe the observed \ce{H_2O} emission, we follow the approach by \citet{RomeroMirzaEA24}: 50 slab models are used to sample profiles in both temperature and column density as a function of the emitting radius. As opposed to \citet{RomeroMirzaEA24}, we first determine the radius as a function of temperature,
\begin{align} \label{eq:RafT}
    R(T) = R_\mathrm{in}\left(\frac{T}{1500\mathrm{\ K}}\right)^{-1/q}.
\end{align}
This relation ensures that the inner radius ($R_\mathrm{in}$ (in au), as listed in Table \ref{tab:Sample}) of our slab models is located at the estimated dust sublimation radius at 1500 K. Using this equation, we sample the annular emitting regions of our 50 slab models for temperatures between 1500 K (the dust sublimation radius; \citealt{Barvainis87}) and 150 K (the condensation temperature of \ce{H_2O}; \citealt{CollingsEA24}) in $\log_{10}$-space. The dust sublimation radius has been calculated following the approach of \citet{DullemondEA01}, $R_\mathrm{in}\sim0.07\sqrt{\left(L_*/L_\odot\right)}$. Both the calculated inner radii and adopted stellar luminosities can be found in Table \ref{tab:Sample}. By simply rewriting Equation \ref{eq:RafT} we obtain a relation for the temperature as a function of the radius,
\begin{align} \label{eq:TafR}
    T(R) = 1500\ \mathrm{K}\left(\frac{R}{R_\mathrm{in}}\right)^{-q}.
\end{align}
\indent For the column density, we use the same profiles as \citet{RomeroMirzaEA24}: a power law (profile I, Equation \ref{eq:N-PL}) and an exponentially tapered power law (profile II, Equation \ref{eq:N-PL-EXT}). In addition, we also try different profiles. Figure \ref{fig:Schematic} provides a schematic overview of the different profiles and their potential behaviours. In particular, we assume a constant number of molecules ($\mathcal{N}_\mathrm{mol}$) with radius, that is each annulus contains the same number of molecules, which automatically resembles a power law with a power of -2 (the emitting area, $A$, is proportional to the square of the radius). We then allow this number of molecules to experience a jump, i.e., the number of molecules gets enhanced (or depleted) by a factor $F_\mathrm{scale}$, at temperatures below a given jump temperature, $T_\mathrm{jump}$ (see Equation \ref{eq:N-NMol}). We have tried profiles with a fixed jump temperature, at 400 K (profile III), and we have kept $T_\mathrm{jump}$ as a free parameter (profile IV). The value of 400 K resembles approximately the boundary temperature between the sublimated (colder temperatures) and gas-phase formed (hotter temperatures) \ce{H_2O} vapour reservoirs (see also \citealt{RomeroMirzaEA24}). The jump temperature can move to higher temperatures, indicating a potential depletion of the hot \ce{H_2O} reservoir, or to colder temperatures, representing a potential enhancement of the cold \ce{H_2O} reservoirs. A parabola provides a smoother description of this behaviour: the jump moving to higher temperatures is captured by an upward parabola (the parabola has a maximum), while a downward parabola (the parabola has a minimum) captures the behaviour of the jump moving to lower temperatures (see also Figure \ref{fig:Schematic}). Therefore, we fitted a parabola in log space (profile V, Equation \ref{eq:N-PILS}). The equations describing the column density profiles of each of these cases are:
\begin{align}
    N_\mathrm{I}(R) & = N_0\left(\frac{R}{1\mathrm{ au}}\right)^{-p}. \label{eq:N-PL} \\
    N_\mathrm{II}(R) & = N_0\left(\frac{R}{1\mathrm{ au}}\right)^{-p}\exp\left[-\left(\frac{R}{R_\mathrm{c}}\right)^\phi\right]. \label{eq:N-PL-EXT} \\
    N_\mathrm{III,IV}(R) & \propto \begin{cases}
                        \frac{\mathcal{N}_\mathrm{mol}}{A} & \mathrm{if }\ T>T_\mathrm{jump}. \\
                        \frac{\mathcal{N}_\mathrm{mol}\times F_\mathrm{scale}}{A} & \mathrm{if }\ T\leq T_\mathrm{jump}.
                   \end{cases} \label{eq:N-NMol} \\
    N_\mathrm{V}(R) & = N_0\left(\frac{R}{1\mathrm{ au}}\right)^{\alpha+\beta\ln\left(R/1\mathrm{ au}\right)}. \label{eq:N-PILS}
\end{align}
\indent The emitting area of each slab model is accounted for in the exact same manner as reported in \citet{RomeroMirzaEA24}. The area is multiplied by the cosine of the disk's inclination ($i$, reported in Table \ref{tab:Sample}), $A=\pi\left(R_2^2-R_1^2\right)\times\cos\left(i\right)$, where $R_1$ and $R_2$ are, respectively, the inner and outer radii of each annulus (see also Equation \ref{eq:RafT}). \\
\indent We fit the profiles, sampled by the 50 slab models, using a Markov Chain Monte Carlo (MCMC) approach, implemented through the \textsc{emcee} python-package \citep{EMCEE}. From the posterior distribution of the MCMC, we take the median values as the best-fit values for the fitted parameter. Additionally, the 18$^\mathrm{th}$ and 84$^\mathrm{th}$ percentiles are taken as the lower and upper 1$\sigma$ uncertainties. We use a reduced-$\chi^2$ as the likelihood function of the MCMC, in order to compare the full model (the sum of the 50 slab models) with the observations. The likelihood has, therefore, the following form:
\begin{align} \label{eq:Chi2}
    \chi^2_\mathrm{red} = \frac{1}{N_\mathrm{data}}\sum_i\frac{w_i\left(F_{\mathrm{obs},i}-F_{\mathrm{model},i}\right)^2}{\sigma_i^2},
\end{align}
where $F_\mathrm{obs}$ is the observed spectrum and $N_\mathrm{data}$ is the number of data points within the used fit regions. $\sigma$ denotes the noise of the observations, determined in a similar manner as in \citet{TemminkEA24}, where the \textit{James Webb Exposure Time Calculator}\footnote{Exposure Time Calculator: \url{https://jwst.etc.stsci.edu/}} has been used to obtain the continuum signal-to-noise (S/N) at the midpoint of each subband given the observational setup. The values for the S/N-ratios are given in Table \ref{tab:S/N-Ratio}. As the noise is significantly higher in Channel 4 and many important \ce{H_2O} lines can be found at these wavelengths, we slightly offset the large noise by introducing weights in the $\chi^2$-formula, $w_i$. All fit regions beyond 20 $\mathrm{\mu}$m have been given a higher weight of 5, while select regions with important transitions probing the cold reservoir have been given a weight of 10 or 15 (for the strongest transitions). While the values are arbitrary, these weights ensure that we are able to properly fit the full rotational spectrum, without favouring the hot reservoir as the transitions at smaller wavelengths have lower flux uncertainties. Fits without any weights favour the hot reservoir. Therefore, the chosen weights ensure that our models properly fit all reservoirs, including the cold one that is best probed by the longer wavelengths that are more noisy given the lower sensitivity of JWST/MIRI's Channel 4. While the choice for the weights is arbitrary, using lower weights will still favour the hotter reservoirs and result in a worse fit for the colder reservoir, whereas higher weights may start overrepresenting the cold reservoir and result in worse fits for the hot one. \\
\indent In the MCMC, we use 20 walkers per free parameter and allow the MCMC to explore the prior space for 50,000 iterations. Table \ref{tab:Priors} lists the prior spaces for each profile. Common priors that are kept the same in each fit are only listed once, for example, the number of molecules, $\mathcal{N}_\mathrm{mol}$, in the jump abundances. For some sources, we run a second fit with a more limited prior space. In these cases, a second, less favourable solution was often found and a number of walkers got stuck in this local minimum, not allowing the fit to converge. The updated prior space was subsequently chosen to exclude these local minima. All fits use selected isolated \ce{H_2O} lines and ortho-para line pairs, which all have been taken from \citet{BanzattiEA24} (see their Figure 3 and Tables 5 and 6). The used fit regions, which are also shifted by the heliocentric velocity, are listed in Table \ref{tab:FitRegs}. We have used fewer regions for CX~Tau and DN~Tau as the \ce{CO_2} $P$- and $R$- branches strongly contribute to their observed emission.
\begin{table}[ht]
    \centering
    \caption{Priors used for the profiles listed in Section \ref{sec:AnalH2O}.}
    \begin{tabular}{c c c}
        \hline\hline
         Profile & Parameter & Priors \\
         \hline
         Temperature-radius & q & $\mathcal{U}\left(0,5\right)$ \\
         \hline
         Power law (I) & $\log_{10}\left(N_0\right)$ & $\mathcal{U}\left(12,30\right)$ \\
                   & $p$ & $\mathcal{U}\left(-10,10\right)$ \\ 
         Exponential taper (II) & $\log_{10}\left(R_\mathrm{c}\right)$ & $\mathcal{U}\left(\log_{10}\left(R_\mathrm{in}\right),1\right)$ \\
                           & $\phi$ & $\mathcal{U}\left(0.5,5\right)$ \\
         \hline
         Jump abundance (III) & $\log_{10}\left(\mathcal{N}_\mathrm{mol}\right)$ & $\mathcal{U}\left(30,50\right)$ \\
                        & $\log_{10}\left(F_\mathrm{scale}\right)$ & $\mathcal{U}\left(-3,5\right)$ \\
         Free temperature (IV) & $T_\mathrm{jump}$ & $\mathcal{U}\left(150,1500\right)$ \\
         \hline
         Parabola (V) & $\log_{10}\left(N_0\right)$ & $\mathcal{U}\left(12,30\right)$ \\
                  & $\alpha$ & $\mathcal{U}\left(-100,100\right)$ \\
                  & $\beta$ & $\mathcal{U}\left(-50,50\right)$ \\
         \hline
    \end{tabular}
    \label{tab:Priors}
\end{table}

\subsection{Rotational \ce{H_2O} spectrum: multiple components}
As a comparison, we also follow \citet{TemminkEA24b} to investigate the pure rotational \ce{H_2O} spectrum using two (in the case of CX~Tau and DN~Tau) or three slab model components. Through trial and error (i.e., by fitting three or two components), we found that only two components were needed to fully describe the rotational \ce{H_2O} spectra of CX~Tau and DN~Tau. This is very likely due to the \ce{H_2O} emission being much weaker compared to the other sources (see Figure \ref{fig:Sample}). We do not use all three approaches presented in \citet{TemminkEA24b}, but only the simplest one, Approach I, which assumes a simple radial gradient. The line optical depths are sufficiently high that the vertical gradient in Approach III does not matter and the results resemble that of Approach I. \\
\indent To agree with the parametric analyses discussed in Section \ref{sec:AnalH2O}, we slightly adapt Approach I: instead of having a circular area surrounded by two annuli, we use three annuli with the inner radius of the first set to the dust sublimation radius, $R_\mathrm{in}$. The other two or three radii are kept free, together with their respective temperatures and column densities. Therefore, Equation 2 of \citet{TemminkEA24b} becomes:
\begin{align}
    F_\mathrm{total} &  = F_1\pi\left[\left(\frac{R_1}{1\mathrm{ au}}\right)^2-\left(\frac{R_\mathrm{in}}{1\mathrm{ au}}\right)^2\right]\cos\left(i\right) \nonumber \\
    & + F_2\pi\left[\left(\frac{R_2}{1\mathrm{ au}}\right)^2-\left(\frac{R_1}{1\mathrm{ au}}\right)^2\right]\cos\left(i\right) \nonumber \\
    & + F_3\pi\left[\left(\frac{R_3}{1\mathrm{ au}}\right)^2-\left(\frac{R_2}{1\mathrm{ au}}\right)^2\right]\cos\left(i\right)
\end{align}
For the sources for which we only fit two components, the third term in the above equation can be ignored. \\
\indent The fits are made using the same MCMC setup, with the same reduced-$\chi^2$ formula taken as the likelihood (see Equation \ref{eq:Chi2}), as in Section \ref{sec:AnalH2O}. The same fit regions are used for the most optimal comparison between the results.

\subsection{Line flux ratios}
Aside from analysing the spectrum through slab models, we follow \citet{BanzattiEA24} and analyse their suggested line flux ratios: $F_{1500\mathrm{K}/3600\mathrm{K}}$ and $F_{3600\mathrm{K}/6000\mathrm{K}}$ (see also the rightmost panel in their Figure 10). In particular, these line ratios will provide information on, respectively, the cold and hot \ce{H_2O} vapour reservoirs. The temperatures indicate the approximate upper level energy. The 1500 K flux is comprised of two isolated \ce{H_2O} transitions, with upper level energies of, respectively, 1448 K (at $\lambda$=23.81676 $\mathrm{\mu}$m) and 1615 K (at $\lambda$=23.89518 $\mathrm{\mu}$m), while the 3600 K and 6000 K fluxes are taken to be the isolated lines at 17.50436 $\mathrm{\mu}$m and 17.32395 $\mathrm{\mu}$m. Given their upper level energies, these lines probe the strength of the different temperature reservoirs and the ratios provide insights into the potential enhancement of the cold reservoir. Following \citet{BanzattiEA24}, we take the sum from the line fluxes of the 1448 K and 1615 K transitions as the total value for $F_{1500\mathrm{K}}$. To determine the line fluxes, we followed the approach of \citet{BanzattiEA12} and \citet{TemminkEA24}, based on the techniques provided by \citet{PascucciEA08}, \citet{NajitaEA10}, and \citet{PontoppidanEA10}: we obtained a Gaussian distribution of measured line flux, using 1000 iterations of adding normally distributed noise (following the $S/N$-ratios in Table \ref{tab:S/N-Ratio}), by fitting Gaussians (using the python-package \textsc{lmfit}, \citealt{lmfit}) to the lines. From these distributions, we take the median value as the line flux and the full width at half maximum (FWHM) as the uncertainty. 

\section{Results} \label{sec:Res}
In the following section, we highlight the results from our different fitting methods. In Section \ref{sec:ResPar}, we present the parametric results for the different fit profiles, while the results of the multiple component analysis are shown in Section \ref{sec:ResMC}. The calculated line fluxes for the different line tracers are displayed in Section \ref{sec:LF}.

\subsection{Parametric analysis} \label{sec:ResPar}
\begin{figure*}[ht!]
    \centering
    \includegraphics[width=0.92\textwidth]{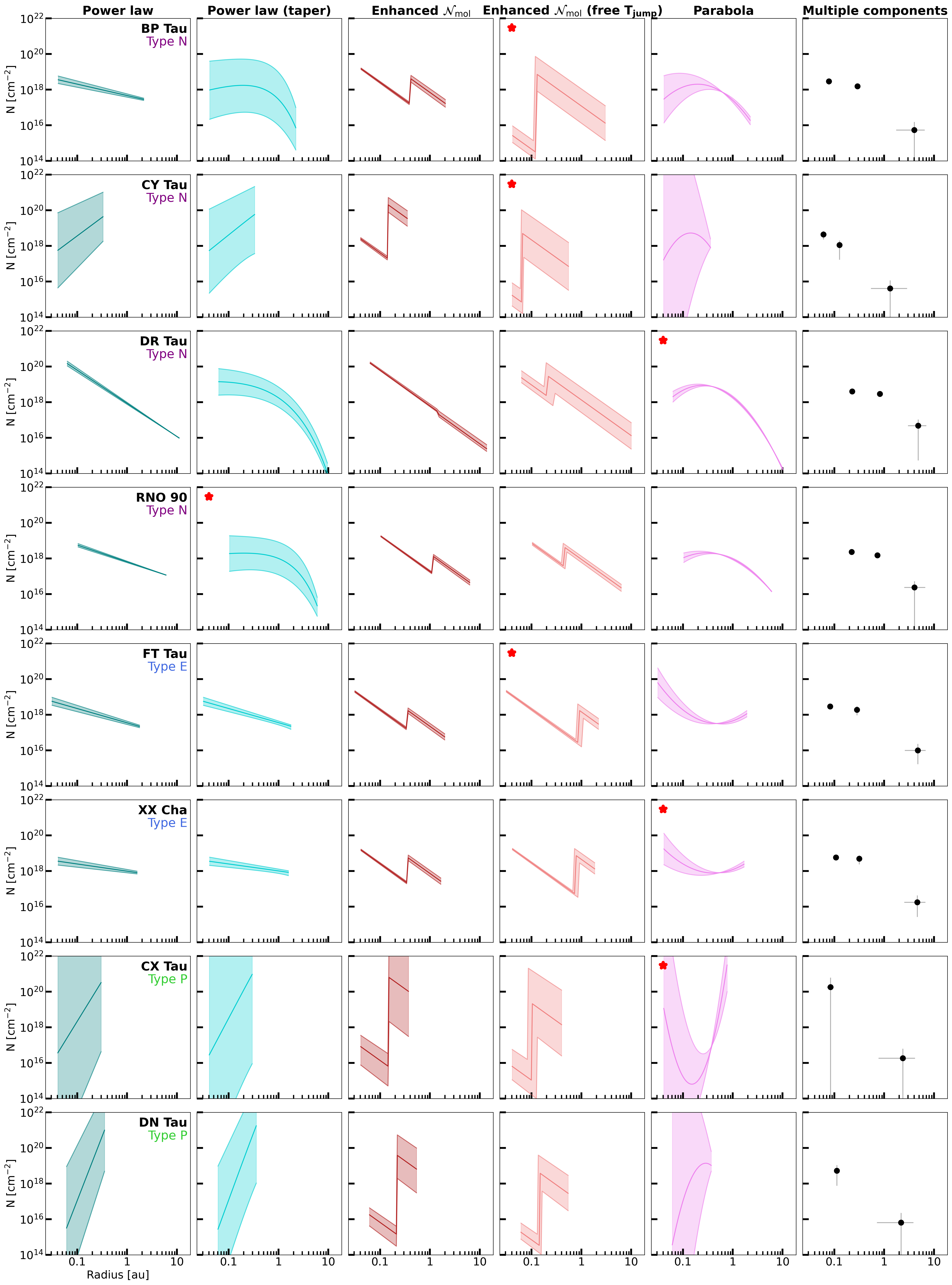}
    \caption{Profiles and multiple components (rightmost panel) fitted for our sample of compact disks using a line width of 4.71 km~s$^{-1}$. The shaded area is the 1$\sigma$-confidence interval, given the uncertainties listed in Tables \ref{tab:ProfileParams-4.71}. The red stars in the top-left corner indicate the best fitting profiles for each disk (see Table \ref{tab:ResultsSum}). We note that the best-fitting profile for DN~Tau is highlighted in Figure \ref{fig:AllProfs-Quad}.}
    \label{fig:AllProfs-4.71}
\end{figure*}
Figure \ref{fig:AllProfs-4.71} displays all fitted profiles for our sample of compact disks with a line width of 4.71 km~s$^{-1}$. Those for the quadrature line widths are given in Figure \ref{fig:AllProfs-Quad}. Given the profiles, we find three groups of disks: BP~Tau, CY~Tau, DR~Tau, and RNO~90 all behave very similarly, which is most notable in the exponentially tapered power laws (second column), the jump abundances with $T_\mathrm{jump}$ at higher temperatures ($T_\mathrm{jump}\gtrsim650$ K, fourth column), and the resulting upward parabolas (fifth column). The resulting profiles for FT~Tau and XX~Cha are also nearly identical, but differ from the other four sources. The jump abundance occurs at a low temperature of $T_\mathrm{jump}\leq$250 K for these disks, while their power laws are very slowly declining and their downward parabolas are very shallow. Finally, aside from their very similar spectra (see Figure \ref{fig:Sample}), the parametrisations of CX~Tau and DN~Tau also behave very similarly to each other, but different from the other disks. Their jump abundance occurs at the intermediate temperatures $T_\mathrm{jump}\sim$500-600 K, while their power laws are monotonically increasing with radius and their parabolas are turned downwards.\\ 
\indent The $\chi^2_\mathrm{red}$ values are listed in Table \ref{tab:ResultsSum}, where we have highlighted the parametrisation with the lowest $\chi^2_\mathrm{red}$-value in bold face. Figures \ref{fig:BPTau-ProfFit}-\ref{fig:XXCha-ProfFit} compare these best-fitting models with the observations and the fitting parameters are also listed in Tables \ref{tab:ProfileParams-4.71} and \ref{tab:ProfileParams-Quad}. These tables also include uncertainty values on the retrieved fitting parameters. Those uncertainties are taken as the 16$^\mathrm{th}$ and 84$^\mathrm{th}$ percentiles of the posterior distributions and, therefore, represent the 1$\sigma$ upper and lower limits, respectively. From the $\chi^2_\mathrm{red}$ values, it is clear that not one single parametrisation provides the best description of all observations. Instead, a combination of the different profiles is preferred. We also note that the $\chi^2_\mathrm{red}$ values of the different parametrisations are overall very similar. \\
\indent Finally, we briefly highlight the resulting temperature slopes (values for $q$, see Equations \ref{eq:RafT} and \ref{eq:TafR}) according to our retrieved profiles. From Tables \ref{tab:ProfileParams-4.71} and \ref{tab:ProfileParams-Quad} it is clear that our retrieved values of $q$ are between 0.35 and 1.30. Overall, we find the lower values for BP~Tau, DR~Tau, FT~Tau, RNO~90, and XX~Cha, while the higher values are found for CX~Tau, CY~Tau, and DN~Tau. The average of the best-fitting parametrisations for all disks suggest a power law slope of $q\sim$0.57 for the temperature, which is somewhat lower than the value found by \citet{RomeroMirzaEA24} of $q\sim$0.69 and the value suggested by thermochemical modelling of $q\sim$0.7 in the surface layers of disks (\citealt{BosmanEA22a}, Vlasblom et al. in prep.). Our average value would be even lower if we had only accounted for the results from the quadrature line widths.

\subsection{Multiple component analysis} \label{sec:ResMC}
Overall, we find rather good agreement between the best-fitting profiles and the multiple components (3 or 2), as can be seen in Figures \ref{fig:AllProfs-4.71} and \ref{fig:AllProfs-Quad}, where the discrete data points indicate the results of such fits. The resulting values are given in Table \ref{tab:MC-Results}, while the $\chi^2_\mathrm{red}$ values are also listed in Table \ref{tab:ResultsSum}. We note that the radial locations shown in Figures \ref{fig:AllProfs-4.71} and \ref{fig:AllProfs-Quad} are not the values for $R_1$, $R_2$, and $R_3$, the outer edge of each emitting area, but these are the central values of each emitting area. The resulting spectra for the best-fitting models are also included in Figures \ref{fig:BPTau-ProfFit}-\ref{fig:XXCha-ProfFit} (red profiles). \\
\indent The results for the multiple component analysis are rather similar between the different disks with either 3 or 2 component fitted: for the 3 component fits, we find that the temperatures for the first component all fall in the 765-970 K range, while those for the second and third components fall in the respective ranges of 355-530 K and 200-300 K. We note that the temperatures of the second and third components for FT~Tau and XX~Cha ($T_2\sim$385 and $T_3<$215) are lower than those for BP~Tau, CY~Tau, DR~Tau, and RNO~90, further suggesting a difference in behaviour as seen for the parametric analysis in Section \ref{sec:ResPar}. The temperatures for CX~Tau and DN~Tau, where only two components were fitted, are also very similar. The first component has a temperature between $\sim$400-500 K for both sources with slightly higher temperatures for DN~Tau, while the second component has a temperature in the range of 190-240 K. \\
\indent Similarly, the column densities follow the same structure, where generally the first component has the highest column density, followed by the second and the third (where applicable). We note, however, that the retrieved column densities cannot be assumed to lie on a simple power law: as can be seen in Figure \ref{fig:AllProfs-4.71} and \ref{fig:AllProfs-Quad} the other parametrisations provide better agreement with the multiple component fits. For example, the downward parabola structure of CX~Tau and DN~Tau cannot be captured by a power law through the individual components. Finally, the high values for the emitting areas of the third component very likely indicate optically thin emission (see also \citealt{TemminkEA24b}). 

\subsection{Line flux ratios} \label{sec:LF}
\begin{figure}[ht!]
    \centering
    \includegraphics[width=\columnwidth]{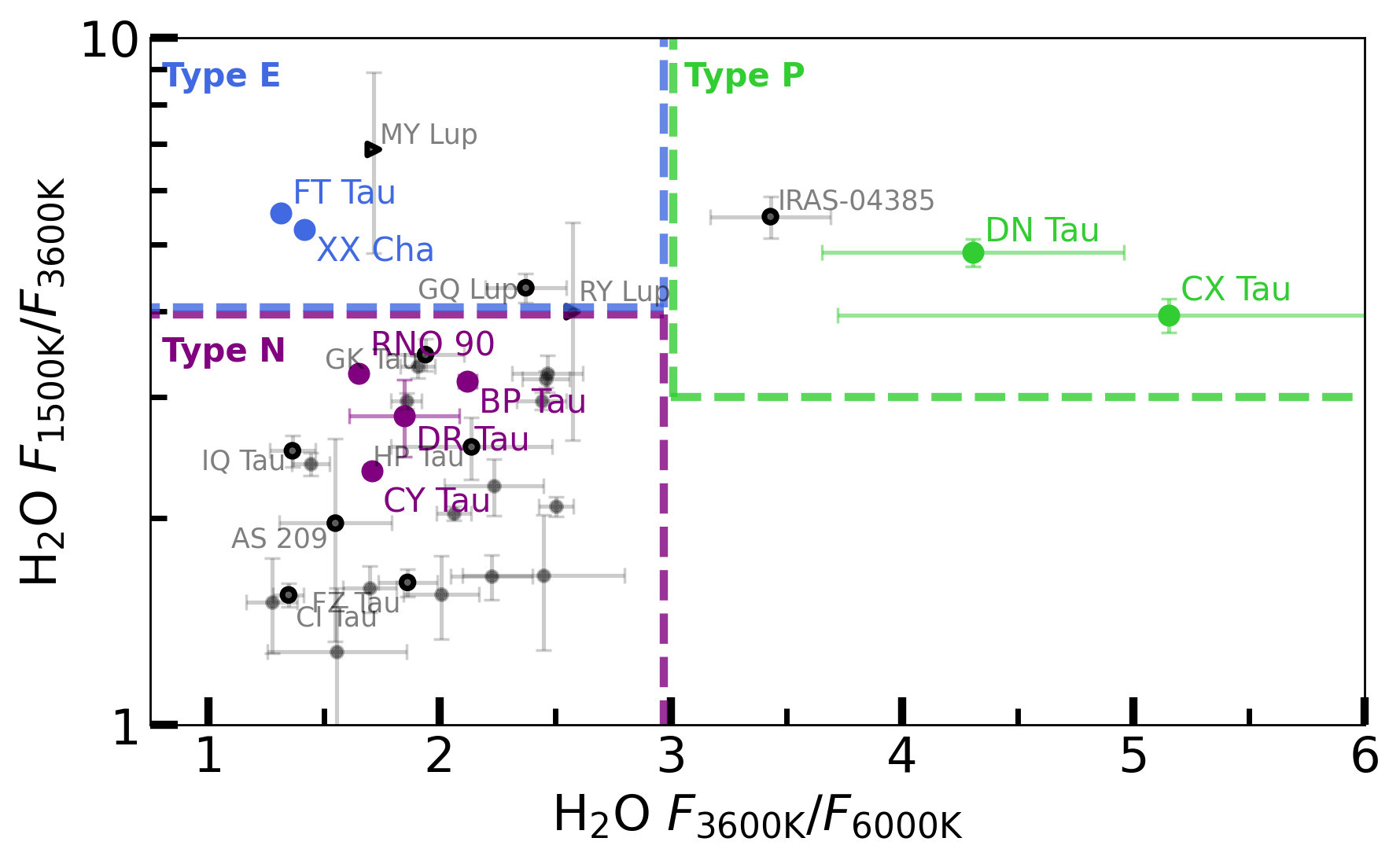}
    \caption{Ratios of the 3600 K/6000 K and 1500 K/3600 K line fluxes, used to investigate the respective strength of each \ce{H_2O} reservoir. The grey data points are adapted from \citet{BanzattiEA24} (see their Table 3), where we have highlighted those analysed by \citet{RomeroMirzaEA24} and potential Type P sources (i.e., MY Lup, RY Lup and IRAS-04385). The coloured data points indicate the line ratios of our sample of compact disks. See Section \ref{sec:Types} for a discussion on the different types.}
    \label{fig:RatioPlot}
\end{figure}

The values for the derived line fluxes are given in Table \ref{tab:LineFluxes}, while the ratios are displayed in Figure \ref{fig:RatioPlot}, which has been adapted from \citet{BanzattiEA24} (see their Figure 10, right panel) to include our sample of compact disks (the coloured data points). The grey data points are those from \citet{BanzattiEA24}, using the values listed in their Table 3. We find that BP~Tau, CY~Tau, DR~Tau, and RNO~90 show similar line flux ratios as the sample of \citet{BanzattiEA24}, while the line fluxes of the cold lines ($F_\mathrm{1500\ K}$) are much stronger in FT~Tau and XX~Cha. CX~Tau and DN~Tau also appear to have higher cold line fluxes, but we note that their hot \ce{H_2O} line fluxes ($F_\mathrm{6000 K}$) are significantly lower. The implications of the locations of the different disks in Figure \ref{fig:RatioPlot} are further discussed in Section \ref{sec:Types}.

\section{Discussion} \label{sec:Disc}
In the following sections, we discuss and interpret our results. From our analysis, it immediately becomes clear that these disks, even though they are all classified as compact, show different behaviours given their \ce{H_2O} spectra and inferred \ce{H_2O} distributions. Not all of these disks show a strong enhanced cold \ce{H_2O} reservoir, suggesting that additional factors are at play within these systems. Within our sample of compact disks, many of them may behave just like large and structured disks, while only a limited number actually show cold \ce{H_2O} emission to be strongly enhanced, likely due to radial drift.

\subsection{Profile types and \ce{H_2O} reservoirs} \label{sec:Types}
\begin{figure*}[ht!]
    \centering
    \includegraphics[width=0.95\textwidth]{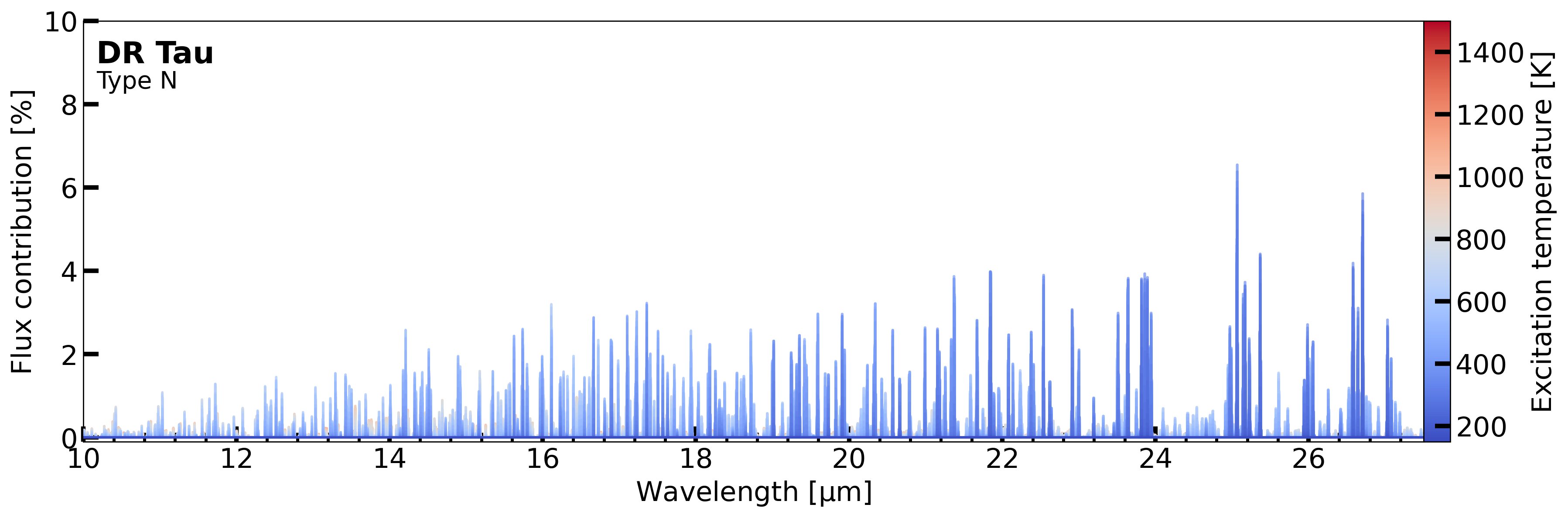}
    \includegraphics[width=0.95\textwidth]{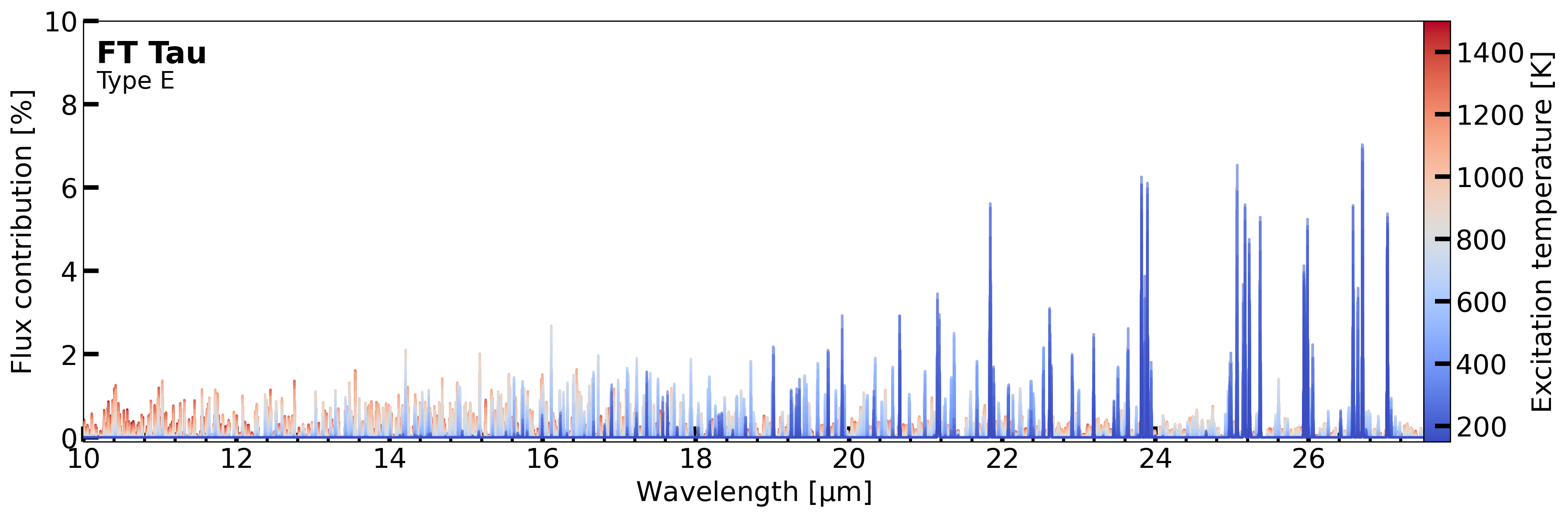}
    \includegraphics[width=0.95\textwidth]{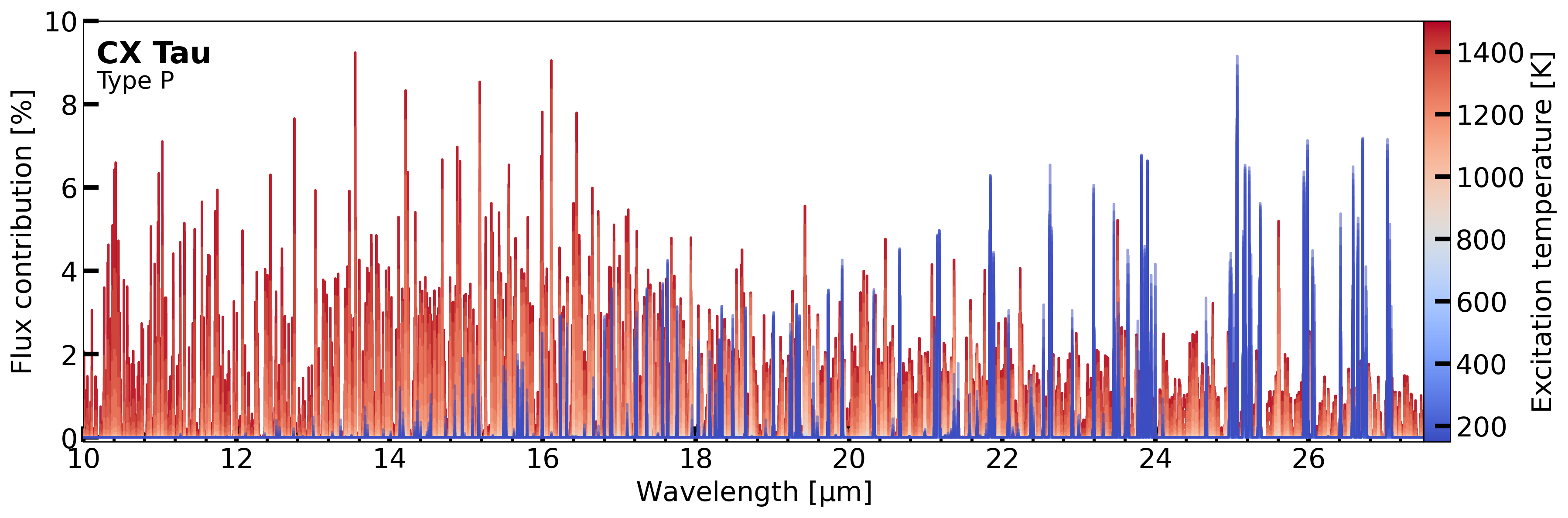}
    \caption{Contributions of each individual slab model with the colour representing the excitation temperature of the respective slab. The contributions are shown for the best fitting parametric models of DR~Tau (Type N), FT~Tau (Type E), and CX~Tau (Type P). The contributions are taken as the percentage with respect to the maximum line flux in the full model. We note that the maximum line flux of CX~Tau is at least a factor of 5 lower than that of the other two types.}
    \label{fig:Contributions1}
\end{figure*}

Given the different observed behaviours (see Section \ref{sec:ResPar}) and the observed line fluxes, we propose to categorise planet-forming disks in at least three different types, based on their \ce{H_2O} reservoirs: Type N, Type E, and Type P (as already highlighted in Figures \ref{fig:Sample} and \ref{fig:RatioPlot}). Type N includes the ``Normal'' disks, which show all three \ce{H_2O} reservoirs (hot with $T>$800 K, intermediate with 400$<T<$800 K, and cold with $T<$400 K). The cold reservoir is present following the temperature gradient, expected to be present in all disks, but not strongly enhanced suggesting that radial drift is not efficiently replenishing the inner disk with icy pebbles. These disks can be found in the lower-left corner of Figure \ref{fig:RatioPlot}, where both ratios ($F_\mathrm{1500K}/F_\mathrm{3600K}$ and $F_\mathrm{3600K}/F_\mathrm{6000K}$) have low values. We expect many of the large and structured disks to be part of this category. \\
\indent The Type E disks include those with very strong cold \ce{H_2O} gas reservoirs, which appear to be significantly enhanced, suggesting efficient radial drift. These disks can be found in the top left corner of Figure \ref{fig:RatioPlot}, indicating their large cold \ce{H_2O} reservoirs by large values for $F_\mathrm{1500K}/F_\mathrm{3600K}$, while their values for $F_\mathrm{3600K}/F_\mathrm{6000K}$ are similar to the Type N disks.  \\
\indent Finally, Type P indicates the \ce{H_2O}-poor (or \ce{CO_2}-rich; \citealt{PontoppidanEA10,VlasblomEA24b}) disks, whose spectra are generally devoid of \ce{H_2O}, but still show strong emission from the cold reservoir. Emission from the other two components is often found to be weak and the emission present can be fitted with either a hot or an intermediate temperature. The lack of \ce{H_2O} emission from both components (i.e., strong emission from both hot and intermediate) could potentially be due to a small inner cavity depleting the reservoirs in these components \citep{GrantEA23,VlasblomEA24}. The cold component could still be prevalent due to radial drift, especially since snowlines are pushed further out in disks with cavities, holding for both the inner and outer regions \citep{TemminkEA23,VlasblomEA24}. These \ce{H_2O}-poor disks can be found in the top right corner of Figure \ref{fig:RatioPlot}, with high values for both $F_\mathrm{1500K}/F_\mathrm{3600K}$ and $F_\mathrm{3600K}/F_\mathrm{6000K}$. We note that the boundaries for the different types in Figure \ref{fig:RatioPlot} are arbitrary. To empirically determine the boundaries between the different types, a larger number of disks needs to be consistently analysed. We leave this for future work (Temmink et al. in prep.). \\
\indent Based on our analysis of the 8 compact disks in the MINDS sample, we suggest that BP~Tau, CY~Tau, DR~Tau, and RNO~90 can all be considered as Type N disks, while FT~Tau and XX~Cha are the only two compact disks with very strong cold \ce{H_2O} reservoirs: we consider them to be Type E disks. CX~Tau and DN~Tau are considered to be Type P disks. Using this categorisation we suggest that the Type N disks can best be described (that is, lowest $\chi^2_\mathrm{red}$-values) by profiles with jump abundances at higher temperatures or upward parabola, which also resembles the power law profile with a strong exponential taper. The Type E disks are the ones with a jump abundance occurring at low temperatures ($T<$250 K), have slow monotonically decreasing power laws, and may be described by a shallow downward parabola. On the other hand, the Type P disks likely have rising power law profiles or jump abundances occurring at the intermediate temperatures. Furthermore, their parabolas have a strong minimum.  \\
\indent The distinction between the different groups also becomes visible when investigating the contribution (with respect to the maximum flux in the observed spectrum) of the 50 individual slab models within the different profiles. As we are fitting 50 different slab models, if they all contributed equally, they would each reach a maximum contribution of 2\% at every wavelength. Figure \ref{fig:Contributions1} displays the flux contribution of the slab models for DR~Tau (top panel, Type N), FT~Tau (middle panel, Type E), and CX~Tau (bottom panel, Type P). The contributions for DR~Tau show that the hot component is the weakest ($\leq$1.5\%), as expected from upward parabola, while the intermediate and cold reservoirs are overall of similar strength ($\sim$3-4\%) given the increasing emitting areas with decreasing temperature. For FT~Tau, the hot and intermediate reservoirs are of similar strength ($\sim$2\%), while the cold reservoir has the strongest contribution ($\leq$6\%). The contributions for CX~Tau show similar strengths in the hot and cold \ce{H_2O} reservoirs ($\leq$8\%), but a negligible intermediate component, as expected from the downward parabola. While slab models probing the hot and cold reservoirs may have similar maximum contributions, fewer hot \ce{H_2O} slabs contribute this much to the spectrum and the spectrum is dominated by the cold reservoir overall. Additionally, we want to emphasise that the best-fitting profile dictates the outcome of Figure \ref{fig:Contributions1}: while the contributions for CX~Tau, because of the parabola, have a strong contribution of the hot component, those of DN~Tau (both Type P disks) will lack the hot component and have significant contributions from the intermediate component as the spectrum is best-fitted by a jump abundance at a temperature of $T_\mathrm{jump}\sim$600 K. Finally, we note that the line-to-continuum ratio of CX~Tau is much lower (factor of $\geq$5) than for the other two sources, highlighting that even the higher percental contributions may not be as noticeable as for the other sources.

\subsection{Role of substructures}
With our sample of compact disks now categorised, we highlight that the analysis of the ALMA visibilities of some of our sources have suggested that these rather compact disks may harbour substructures. In particular, gaps and rings have been proposed for the outer regions ($>$15 au) of BP~Tau \citep{JenningsEA22,ZhangEA23,GasmanEA25}, DN~Tau \citep{LongEA18,ZhangEA23}, DR~Tau \citep{JenningsEA22,ZhangEA23,GasmanEA25}, and FT~Tau \citep{LongEA18,JenningsEA22,ZhangEA23}. With substructures hypothesised in disks across all three types, it is clear that the role of substructures in setting the inner disk \ce{H_2O} reservoirs is not yet well understood (see also \citealt{GasmanEA25}). Furthermore, as modelling works have shown that the influence depends on the gap location, the time at which the structure formed, and how leaky the traps are, this implies that substructures may play a very diverse role in setting the inner disk reservoirs \citep{KalyaanEA23,SellekEA24,EasterwoodEA24}. Age may also play a role \citep{MahEA23}, but we note that the majority of our sample consists of sources from the Taurus star-forming region and it can, therefore, be assumed that these disks all have a similar age. Therefore, our sample may be rather homogeneous and a larger sample, consisting of sources from a variety of different star-forming regions and with different ages, is needed to not only investigate the role of the structure formation time, but also the drift timescale \citep{MahEA23,MahEA24}. \\
\indent Our introduced types may, therefore, be comprised of a wide variety of disks. Higher-resolution ALMA observations (spatial resolution of $<$0.1", but preferably as high as possible) and observations with infrared observatories are needed to confirm and further study these substructures and their potential roles in setting the inner disk abundances, as well as hunt for substructures at smaller scales.

\subsection{Profile implications} \label{sec:DiscBI}
The following section discusses the differences in more detail. In particular, we further highlight the behaviours of the profiles and what this implies. In Section \ref{sec:ProfComp}, we also discuss how well the different profiles fit for each type. \\
\indent For the Type N disks, we find that all disks prefer a jump abundance at higher temperatures of generally $T_\mathrm{jump}\gtrsim$700 K or a downward parabola (maxima generally at temperatures above 600 K). This behaviour suggests a potential depletion of the hotter component with respect to the stronger intermediate and cold components. This `depletion' could simply mean that the hot \ce{H_2O} reservoir is not as abundant as the intermediate and cold ones and, therefore, does not follow a simple power law profile, but it may also have a more physical explanation. As suggested by theoretical models \citep{SellekEA24,HougeEA25}, the inward drift of icy pebbles not only increases the molecular reservoirs when crossing the snowlines, but also increases the continuum optical depth ($\tau$) and, therefore, the continuum $\tau$=1 layer may be moved to higher layers and obscure the underlying \ce{H_2O} reservoirs. \\
\indent On the other hand, the optical depth of the \ce{H_2O} lines themselves may play an important role, which could reach values of $\tau\sim$3000 \citep{TemminkEA24b}. Therefore, it is possible that only the very top region of the emitting layer can be probed, well above the dust continuum $\tau$=1 layer, while the remainder of the reservoir remains invisible given the high optical depths. As the line optical depth is found to be higher for the hottest component compared to the cooler components \citep{TemminkEA24b}, this may play an important role in setting the observable column density for \ce{H_2O} in the innermost region. \\
\indent Finally, the destruction of \ce{H_2O} could also play a role in depleting the inner regions. One of the main destruction routes in these high atmospheric layers of disks is photodissociation. The photodissociation must, however, occur on timescales faster than the gas-phase formation, which is efficient at $>300$ K \citep{BosmanEA21}, to effectively destroy \ce{H_2O}. Additionally, \ce{H_2O} can be destroyed in these layers through collisions with \ce{C^+}, \ce{Si^+}, and \ce{H^+} \citep{KampEA13}. There is no consensus on which explanation for the flattening of the profiles in the inner regions is correct, as it may be a combination of all three: dust obscuration, optical depth, and destruction. We expect that it is more than likely that different combinations are preferred by different disks and the importance of each explanation will also differ. \\
\indent The Type E disks, FT~Tau and XX~Cha, have most notably the jump abundance at low temperature ($T_\mathrm{jump}<$250 K) or a shallow downward parabola. As drift can be expected to play to a certain extent a role in all disks, both smooth and structured, where the leakiness of the substructures determines the inward flow of material \citep{MahEA23,MahEA24,GasmanEA25}, the cold component should be present in most if not all disks. The strength of the cold component in FT~Tau and XX~Cha, and therefore the jump in abundance at $T_\mathrm{jump}\sim$250 K and the slowly decreasing power law, can be explained by an extreme or very efficient inward flux of the icy pebbles, enhancing the abundance of this cold component near the \ce{H_2O} snowline with respect to the other disks.  \\
\indent We note that not all disks in the Type E category may undergo efficient radial drift. Instead, an accretion outburst that significantly heats up the disk and, therefore, enlarges the emitting area of the cold component may also result in disks falling into this category. In particular, it has been found that XX~Cha is highly variable in its accretion \citep{ClaesEA22}. Even though its variability timescales have not yet been constrained, the observed variation with the X-shooter instrument is on a scale of almost 2 dex (i.e., a factor of 100). Therefore, the strong cold component of XX~Cha may be due to its strong variably accretion instead of efficient drift. Furthermore, recent work attributed the strong cold \ce{H_2O} reservoir observed in the disk of EX~Lup to an accretion outburst and the subsequent sublimation of \ce{H_2O}-ice given the snowline being pushed outwards \citep{SmithEA25}. For FT~Tau, the other disk identified as a Type E disk in our sample, there are currently no claims of accretion outburst or a strongly variable accretion rate. The reported accretion rate for FT Tau gives values between $\log_{10}\left(\dot{M}_\mathrm{acc}\right)\sim$-7.5 and -8.9 \citep{GarufiEA14,GangiEA22} (with $\dot{M}_\mathrm{acc}$ given in $M_\odot$~year$^{-1}$), but have not been monitored closely over time. To confirm the role of accretion outbursts and variability in setting the inner disk chemical reservoirs, the accretion rates need to be monitored over an extended period of time to also infer the variability timescales. Also, if present, a larger sample of these sources needs to be observed with JWST-MIRI/MRS and analysed in a consistent manner. \\
\indent The preferred parametrisations of CX~Tau and DN~Tau, the Type P disks, include a monotonically increasing power law, a jump abundance at the intermediate temperatures, and a downward parabola. These profiles all imply a depletion of the hot or intermediate \ce{H_2O} component with respect to the colder reservoir, which is clearly the strongest. This may suggest a potential dust and gas cavity (see, for example, \citealt{SalykEA25}), where the cold \ce{H_2O} is most prominent due to sublimation of \ce{H_2O} ice near the cavity edge (see also \citealt{VlasblomEA24}). \\
\indent Finally, we note that \citet{BanzattiEA23b} described two compact disks (GK~Tau and HP~Tau) having an ``excess'' in their cold \ce{H_2O} reservoir with respect to a large and structured disk (CI~Tau). Here, we find that similar line flux ratios,  as for their two compact disks and for our Type N disks, do not necessarily imply an enhancement of the cold \ce{H_2O} reservoir, but rather a smooth distribution from hot and/or warm to cold. A true enhancement, as found for our Type E disks, produces an even stronger excess in the cold \ce{H_2O} lines with respect to their warmer/hotter counterparts. This may also imply that CI~Tau is actually depleted in the cold \ce{H_2O} and, therefore, falls to the lower left corner of Figure \ref{fig:RatioPlot}. Therefore, the Type E compact disks of FT~Tau and XX~Cha most closely conform with the hypothesis posed by \citet{BanzattiEA20} and the overall situation appears to be more complex. \\
\indent Overall, we find that the jump abundance with the jump temperature ($T_\mathrm{jump}$) kept free is amongst the favoured fit for all three types. Therefore, we suggest that this profile can be fitted to investigate the \ce{H_2O} reservoirs in all disks. Furthermore, the location of the jump, as discussed for each type above, may suggest to which type the disk belongs.

\subsection{Line width comparison} \label{sec:LWComp}
Not yet addressed are the choices for the different line widths: a fixed value of 4.71 km~s$^{-1}$ or the sum in quadrature of a fixed turbulent line width and the thermal line width, given the temperature of the slab model. Intuitively, the quadrature line profiles may be preferred, as the line width changes accordingly with the temperature. One downside is that the strength of turbulence is not known in the inner regions of these disks, therefore the turbulent component of the line width (fixed to 1 km~s$^{-1}$) may be stronger or weaker. Thus far, a few studies have measured the turbulence in the outer regions of large disks using ALMA, resulting in values of $<0.2$ km~s$^{-1}$ \citep{PCEA24} or up to $\sim$35\% of the sound speed ($<$0.35$c_s$; \citealt{FlahertyEA15,FlahertyEA17,FlahertyEA20}). We find that seven out of eight of our disks prefer the fixed value of 4.71 km~s$^{-1}$. This may suggest that the inner regions of these disks are more turbulent and higher values need to be assumed for the turbulent component ($\Delta V_\mathrm{turb}>$1.0 km~s$^{-1}$). \\
\indent The preferred larger line width may be related to the change in the optical depth of the lines, which decreases with larger line widths ($\tau\sim\ N/\Delta V$, with $\Delta V$ the line width). As the quadrature line widths are overall smaller (see also \citealt{RomeroMirzaEA24}), these models will have larger optical depths. Furthermore, it becomes apparent from the contribution plots (Figures \ref{fig:Contr-DRTau}-\ref{fig:Contr-FTTau}) that the quadrature line profiles have stronger fluxes in certain lines of the cold component (see, for example, transitions around 25.05 or 26.70 $\mathrm{\mu}$m). Given the added weights to the lines fitted at the longest wavelengths ($>$20 $\mathrm{\mu}$m), we expect the fitting method to find the best profile describing the observed spectrum, regardless of the lower sensitivity of the larger wavelengths observed with JWST-MIRI. Therefore, it is possible that the cold component of the majority of our sample is less optically thick compared to the other disks. \\
\indent Finally, we notice that our profiles show no qualitative differences between the different line widths (see Figures \ref{fig:AllProfs-4.71} and \ref{fig:AllProfs-Quad}), except for the parabola of DN~Tau which changes between upward and downward. Therefore, we conclude that the choice of line width does not really matter for the profiles, but note that the optical depth of the reservoirs, as discussed above, does depend on the chosen line width.

\subsection{Comparisons with previous analyses}
\begin{figure}[ht!]
    \centering
    \includegraphics[width=\columnwidth]{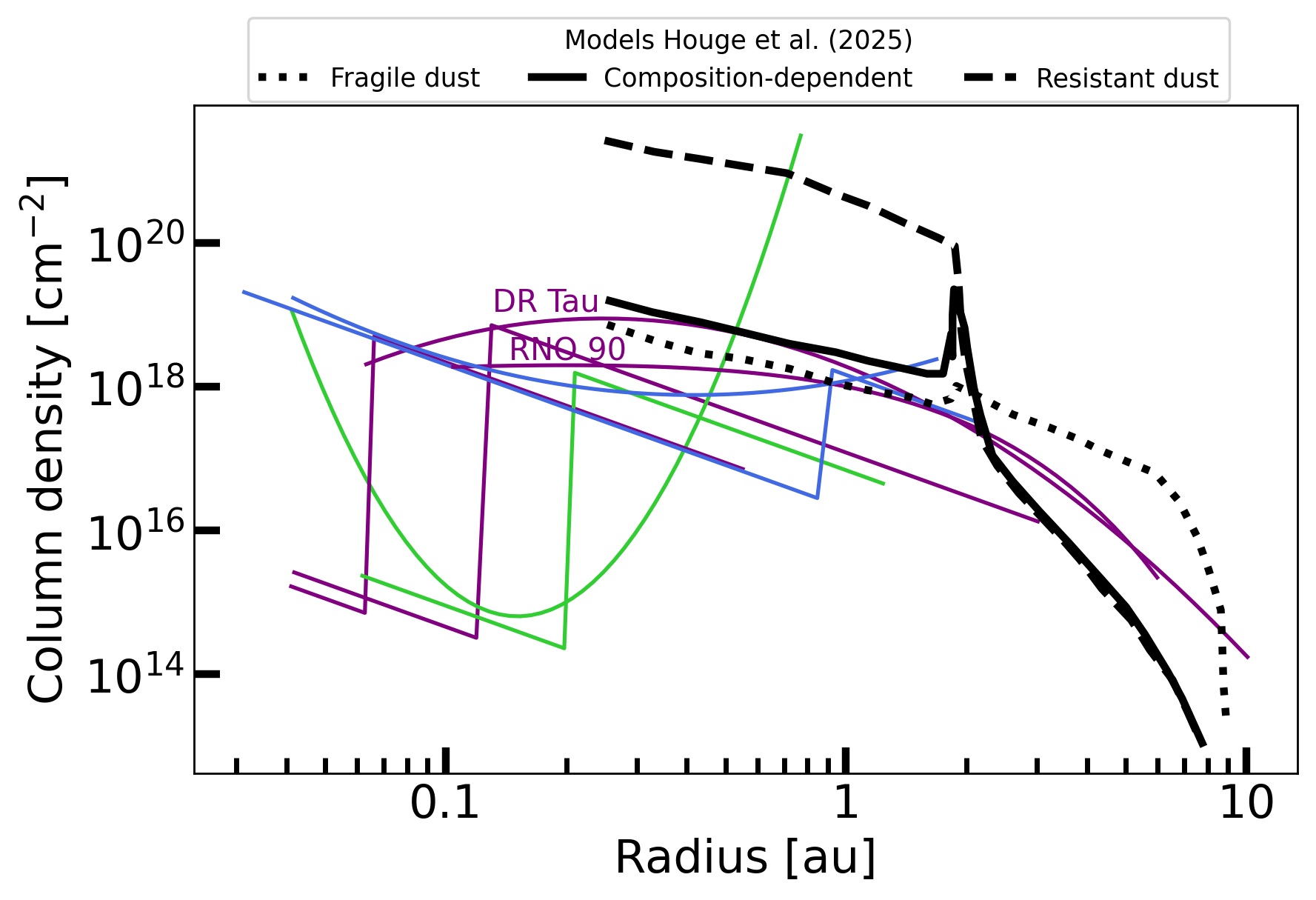}
    \caption{Comparisons between our parametric models with the lowest $\chi_\mathrm{red}^2$-values and the prediction from the radially varying vertical models from \citet{HougeEA25} for all three tested dust types: fragile (dotted), composition-dependent (solid), and resistant (dashed). The profiles of our disks identified to be Type N are shown in purple, while those of the Type E and Type P disks are displayed in, respectively, blue and green.}
    \label{fig:HougeComparison}
\end{figure}

In this section, we compare our analysis with the works of \citet{RomeroMirzaEA24} and \citet{GasmanEA25}, who, respectively, fitted (tapered) power-laws to seven other disks and analysed the \ce{H_2O} emission in ten structured disks. Additionally, we compare our results with the interpretation of the models by \citet{HougeEA25}. \\
\indent \citet{RomeroMirzaEA24} only used power law and tapered power law profiles (our profiles I and II) to fit the column density distributions of four compact disks (FZ~Tau, GK~Tau, GQ~Tau, and HP~Tau) and three larger disks (AS~209, CI~Tau, and IQ~Tau). They find that all these disks can be well described by a simple power law, while both CI~Tau and FZ~Tau prefer the exponentially tapered power law, potentially suggesting a ring-like distribution of the \ce{H_2O} reservoir. All disks are also captured in the analysis of \citet{BanzattiEA24} and are, therefore, displayed in Figure \ref{fig:RatioPlot}, in which we have highlighted their respective positions. All of them, except for maybe GQ~Lup, fall within our selection of Type N disks. GQ~Lup may also be a Type N disk, given that it is located towards the lower right of Figure \ref{fig:RatioPlot} with respect to FT~Tau and XX~Cha and has, therefore, a stronger intermediate component. As stated before, the current definition of the types on Figure \ref{fig:RatioPlot} is somewhat arbitrary and refining the boundaries for each disk type requires a larger sample of sources, which will be addressed in future work (Temmink et al. in prep.). We also cannot exclude the possibility that other behaviours and types may exist outside the parameter space covered by the 8 disks in our sample. As all disks analysed by \citet{RomeroMirzaEA24} may be Type N disks, their profiles can indeed be described by power laws (similar to CY~Tau), but may also be represented by jump abundances at higher temperatures or upward parabolas. \\
\indent \citet{GasmanEA25} compared a sample of 10 structured disks and put their \ce{H_2O} emission in context with respect to DR~Tau. They note that the \ce{H_2O} reservoirs are complicated for these structured disks and that the \ce{H_2O} reservoirs strongly depend on the age of the system, the mechanisms that open a gap in the outer disk, and the leakiness of such substructures (see also \citealt{MahEA24}). We note that BP~Tau and DR~Tau are the two disks overlapping between their sample and ours, highlighting that compact disks may be structured as well. Relative to DR~Tau, there is a distinct group of large structured disks (CI~Tau, DL~Tau, GW~Lup, and V1094~Sco) that have a weaker cold \ce{H_2O} reservoir, which may end up in the lower left corner of Figure \ref{fig:RatioPlot}, where the position of CI~Tau is already highlighted. Another group of structured disks (SY~Cha and Sz~98) shows, on the other hand, a very strong cold component (stronger than DR~Tau) and may, therefore, end up in the upper left corner of Figure \ref{fig:RatioPlot} near FT~Tau and XX~Cha. This suggests that the \ce{H_2O} reservoirs of planet-forming disks are much more complicated and one cannot simply separate them into different types based on their size and/or dust structures, as already discussed in Section \ref{sec:DiscBI}. \\
\indent Finally, we compare our best-fitting profiles to the predicted model column density profiles by \citet{HougeEA25} (for a star with $L_*$=4 L$_\odot$, see Figure \ref{fig:HougeComparison}). Their column density profiles are given for the observable \ce{H_2O} reservoir above the $\tau$=1 layer at 20 $\mathrm{\mu}$m. Their predictions are given for a fixed ratio between the chemical and mixing timescales ($t_\mathrm{chem}/t_\mathrm{mix}$=0.01) or for a fixed chemical timescale ($t_\mathrm{chem}$=10 years) with a radially varying mixing timescale. Furthermore, their models include three types of dust models: fragile ($v_\mathrm{frag}$=1 m~s$^{-1}$), composition-dependent ($v_\mathrm{frag}$=1-10 m~s$^{-1}$), and resistant ($v_\mathrm{frag}$=10 m~s$^{-1}$). They find that for sufficient fragile dust (i.e., in the fragile or composition-dependent cases) the column density is set by the vapour-to-dust mass ratio and the dust opacity, as the profiles for those models do not strongly depend on any disk properties and, therefore, do not evolve over time (see also \citealt{SellekEA24}). While a direct comparison with the models of \citet{HougeEA25} cannot be made given the stellar parameters used in their models, we compare the overall shape of the models with our best-fitting column density profiles in Figure \ref{fig:HougeComparison}. In particular, we make the comparisons with the models with a fixed chemical timescale and radially varying vertical mixing for the different dust types at a time of 1~Myr, an appropriate assumed age for the majority of our sources (see, for example, \citealt{KroliEA21,Luhman23}). \\
\indent Qualitatively, the models and observations agree in the sense that the column density profiles decrease towards larger radii. However, our profiles decrease or flatten off towards the inner regions (see, for example, DR~Tau and RNO~90). This could be due to the fact that the expression used by \citet{HougeEA25} to derive the abundance of water in the surface layers assumes that the chemical timescale (i.e., the destruction timescale) is always much shorter than the mixing timescale. Under their assumptions, the abundance should saturate inside ~0.16 au at a level equal to that of the midplane, resulting in the profile flattening off. This level, namely the limit of $N\sim$10$^{20}$ cm$^{-2}$ explored by \citet{SellekEA24}, would be equivalent to negligible destruction under the assumption that the \ce{H_2O} and dust particles behave the same and on similar timescales (that is, as long as the dust is not depleted on faster timescales than the \ce{H_2O}). Our best-fitting profiles, except for the outer regions of CX~Tau, do not exceed a column density of $N\sim$10$^{20}$ cm$^{-2}$, which is in agreement with the limit proposed by \citet{SellekEA24}. \\
\indent Quantitatively, the retrieved column densities fall below all the model profiles for many of our sources, even for the most conservative model predictions with the fragile dust. On the one hand, while their expected column density profiles are based on the increased dust obscuration following radial drift, the importance of the optical depth of the \ce{H_2O} lines themselves is not explored, which may be important (as discussed in Section \ref{sec:DiscBI}). Alternatively, this may suggest that for our sample of disks the ratio of the chemical and mixing timescales is smaller than their assumptions. This may result from shorter photodissociation timescales which are typically expected to be $<$1 year in the upper most layers of the disk (\citealt{BosmanEA21}, Vlasblom et al. in prep.) and depend source-by-source on the stellar properties (i.e., the UV irradiation field). This may suggest that a chemical timescale of $t_\mathrm{chem}$=10 yr may be too long. Future 2D models should further investigate the physical and chemical processing of \ce{H_2O} in the inner regions and explore how effective the combination of increased dust obscuration following radial drift and line optical depth are.

\subsection{Other molecular species}
While our analysis focuses solely on the \ce{H_2O} emission, the spectra of these compact disks contain more molecular species. In this section, we only highlight the other molecular species detected and leave the analysis of this emission for future work. We start by noting that CX~Tau \citep{VlasblomEA24b} and DR~Tau \citep{TemminkEA24,TemminkEA24b} have dedicated papers analysing their molecular emission and we, therefore, refer the reader to these papers. For the other disks in our sample, the commonly observed species (\ce{OH}, \ce{CO_2}, \ce{HCN}, and \ce{C_2H_2}) are detected in nearly all disks. For some disks, we also report the (potential) detection of \ce{CH_4} in CY~Tau and that of some of the isotopologues, such as \ce{^{13}CCH_2} in CY~Tau and both \ce{^{13}CO_2} and \ce{CO^{18}O} in DN~Tau, which is very similar to CX~Tau \citep{VlasblomEA24b}. The different species detected in each disk are summarised in Table \ref{tab:OMDets}, while Appendix \ref{sec:OMSpecies} contains a more elaborate discussion of the detected species. \\

\begin{table*}[ht!]
    \centering
    \caption{Summary of the other molecules detected in our sample of disks.}
    \begin{tabular}{c c c c c c c c c c}
        \hline\hline
        Source & Type & \ce{OH} & \ce{CO_2} & \ce{^{13}CO_2} & \ce{CO^{18}O} & \ce{HCN} & \ce{C_2H_2} & \ce{^{13}CCH_2} & \ce{CH_4} \\
        \hline
        BP~Tau & N & \checkmark & \checkmark & $\times$ & $\times$ & \checkmark & $\times$ & $\times$ & $\times$ \\
        CX~Tau$^\alpha$ & P & \checkmark & \checkmark & \checkmark & \checkmark(?) & \checkmark & \checkmark & $\times$ & $\times$ \\
        CY~Tau & N & \checkmark & \checkmark & $\times$ & $\times$ & \checkmark & \checkmark & \checkmark & \checkmark(?) \\
        DN~Tau & P & \checkmark & \checkmark & \checkmark & \checkmark(?) & $\times$ & $\times$ & $\times$ & $\times$ \\
        DR~Tau$^\alpha$ & N & \checkmark & \checkmark & $\times$ & $\times$ & \checkmark & \checkmark & $\times$ & $\times$ \\
        FT~Tau & E & \checkmark & \checkmark & $\times$ & $\times$ & \checkmark & \checkmark & $\times$ & $\times$ \\
        RNO~90 & N & \checkmark & \checkmark & $\times$ & $\times$ & \checkmark & \checkmark & $\times$ & $\times$ \\
        XX~Cha & E & \checkmark & \checkmark & $\times$ & $\times$ & \checkmark & \checkmark & $\times$ & $\times$ \\
        \hline
    \end{tabular}
    \label{tab:OMDets}
    \tablefoot{$^\alpha$: see \citet{VlasblomEA24b} for the molecules detected in CX~Tau and \citet{TemminkEA24,TemminkEA24b} for those in DR~Tau.}
\end{table*}

\section{Conclusions and summary} \label{sec:CS}
In this work, we analyse the pure rotational \ce{H_2O} JWST-MIRI/MRS spectra of 8 millimetre-compact dust disks. We expand upon existing techniques to provide detailed parametric profiles of the column densities beyond (exponentially tapered) power laws. This leads us to infer the best-fitting radial profiles and investigate whether these compact disks show signatures of enhanced reservoirs due to radial drift or whether the overall situation is more complex. Our main conclusions are summarised as follows:
\begin{itemize}
    \item The pure rotational \ce{H_2O} spectra of compact dust disks are very different from each other. Half of the disks show similar strengths of the different \ce{H_2O} reservoirs, conforming with large and structured disks, while others show clear signs of an enhanced cold reservoir, even beyond that found in previous comparisons of compact and large disks.
    \item Different combinations of parametrisations can be used to describe the observed reservoirs in different disks, leading to the conclusion that planet-forming disks can be grouped into at least three different types (N, E, and P), based on their mid-infrared \ce{H_2O} spectra.
    \item The parametrisation that assumes a constant of number of molecules with radius (i.e., the column density is proportional to negative the square of the radius) and has a jump (either an enhancement or depletion) in the abundance at a free temperature is able to provide a good fit for the column density profiles for all disks. The temperature of the jump varies per disk but is found to be similar within each of the different types (N, E, or P).
    \item We find that the column density profiles, and therefore the distribution of the \ce{H_2O} reservoirs, are generally best fitted with a fixed line width of $\Delta V$=4.71 km~s$^{-1}$. As the quadrature sum is intuitively and physically more correct, this may suggest that the turbulence is stronger in the inner disk and, therefore, a larger value for the turbulent component needs to be assumed ($\Delta V_\mathrm{turb}>$1 km~s$^{-1}$).
    \item Overall, a good agreement is found between the column densities retrieved through a multiple component analysis and the profiles obtained through the parametric models. The parametric models show that one cannot assume a simple power law through the column densities derived from a multiple component fit, but it is recommended to fit one of the other parametric models used throughout this work.
\end{itemize}
This work has shown that not all millimetre-compact disks have signatures of strongly enhanced cold \ce{H_2O} reservoirs following radial drift. Instead, half of our sample behaves similarly to larger, more structured disks. Therefore, we have introduced a new categorisation, given the behaviour of the \ce{H_2O} reservoirs seen for our disks: Type N, Type E, and Type P disks. Here, N stands for ``Normal" disks, whose spectra have contributions from all components (hot, intermediate, and cold). Type P disks are the \ce{H_2O}-poor disks, whose spectra are dominated by other molecular species, yet have a strong contribution from the cold reservoir. Finally, the Type E disks are the ones that show enhanced cold \ce{H_2O}, that may be the result of the strong inwards drift of icy pebbles or, perhaps, the product of an accretion outburst that pushes out the \ce{H_2O} and increases the emitting area of the cold \ce{H_2O} reservoir. Only two of our 8 compact disks fit into this final category, further highlighting that not all compact disks have enhanced cold \ce{H_2O} reservoirs. These three types provide a new classification of disks, purely based on their rotational \ce{H_2O} emission and may be used to further explore trends involving system properties. Higher resolution ALMA observations and/or observations with infrared interferometers are required to further study the occurrence of substructures in these disks, on both small and large scales, and how they may affect the observable reservoirs.


\begin{acknowledgements}
    The authors would like to thank the referee for many thoughtful, constructive comments that helped improve the manuscript. \\
    \indent This work is based on observations made with the NASA/ESA/CSA James Webb Space Telescope. The data were obtained from the Mikulski Archive for Space Telescopes at the Space Telescope Science Institute, which is operated by the Association of Universities for Research in Astronomy, Inc., under NASA contract NAS 5-03127 for JWST. These observations are associated with program \#1282. The following National and International Funding Agencies funded and supported the MIRI development: NASA; ESA; Belgian Science Policy Office (BELSPO); Centre Nationale d’Etudes Spatiales (CNES); Danish National Space Centre; Deutsches Zentrum fur Luft- und Raumfahrt (DLR); Enterprise Ireland; Ministerio De Econom\'ia y Competividad; Netherlands Research School for Astronomy (NOVA); Netherlands Organisation for Scientific Research (NWO); Science and Technology Facilities Council; Swiss Space Office; Swedish National Space Agency; and UK Space Agency. \\
    \indent M.T., A.D.S., E.F.v.D., and M.V. all acknowledge support from the ERC grant 101019751 MOLDISK. D.G. thanks the Belgian Federal Science Policy Office (BELSPO) for the provision of financial support in the framework of the PRODEX Programme of the European Space Agency (ESA). E.F.v.D. also acknowledges support the Danish National Research Foundation through the Center of Excellence ``InterCat'' (DNRF150). E.F.v.D., I.K., and A.M.A. acknowledge support from grant TOP-1 614.001.751 from the Dutch Research Council (NWO). T.H., K.S. and M.S. acknowledge support from the European Research Council under the Horizon 2020 Framework Program via the ERC Advanced Grant Origins 83 24 28. A.C.G. acknowledges support from PRIN-MUR 2022 20228JPA3A “The path to star and planet formation in the JWST era (PATH)” funded by NextGeneration EU and by INAF-GoG 2022 “NIR-dark Accretion Outbursts in Massive Young stellar objects (NAOMY)” and Large Grant INAF 2022 “YSOs Outflows, Disks and Accretion: towards a global framework for the evolution of planet forming systems (YODA)”. G.P. gratefully acknowledges support from the Carlsberg Foundation, grant CF23-0481 and from the Max Planck Society. B.T. is a Laureate of the Paris Region fellowship program, which is supported by the Ile-de-France Region and has received funding under the Horizon 2020 innovation framework program and Marie Sklodowska-Curie grant agreement No. 945298. \\
    \indent This work also has made use of the following software packags that have not been mentioned in the main text: NumPy, SciPy, Astropy, Matplotlib, pandas, IPython, Jupyter \citep{Numpy,Scipy,AstropyI,AstropyII,AstropyIII,Matplotlib,pandas,IPython,Jupyter}
\end{acknowledgements}

\bibliographystyle{aa}
\bibliography{Bibliography}


\newpage
\onecolumn
\begin{appendix}
\section{Velocity shifted slab models}
\begin{figure}[ht!]
    \centering
    \includegraphics[width=0.5\textwidth]{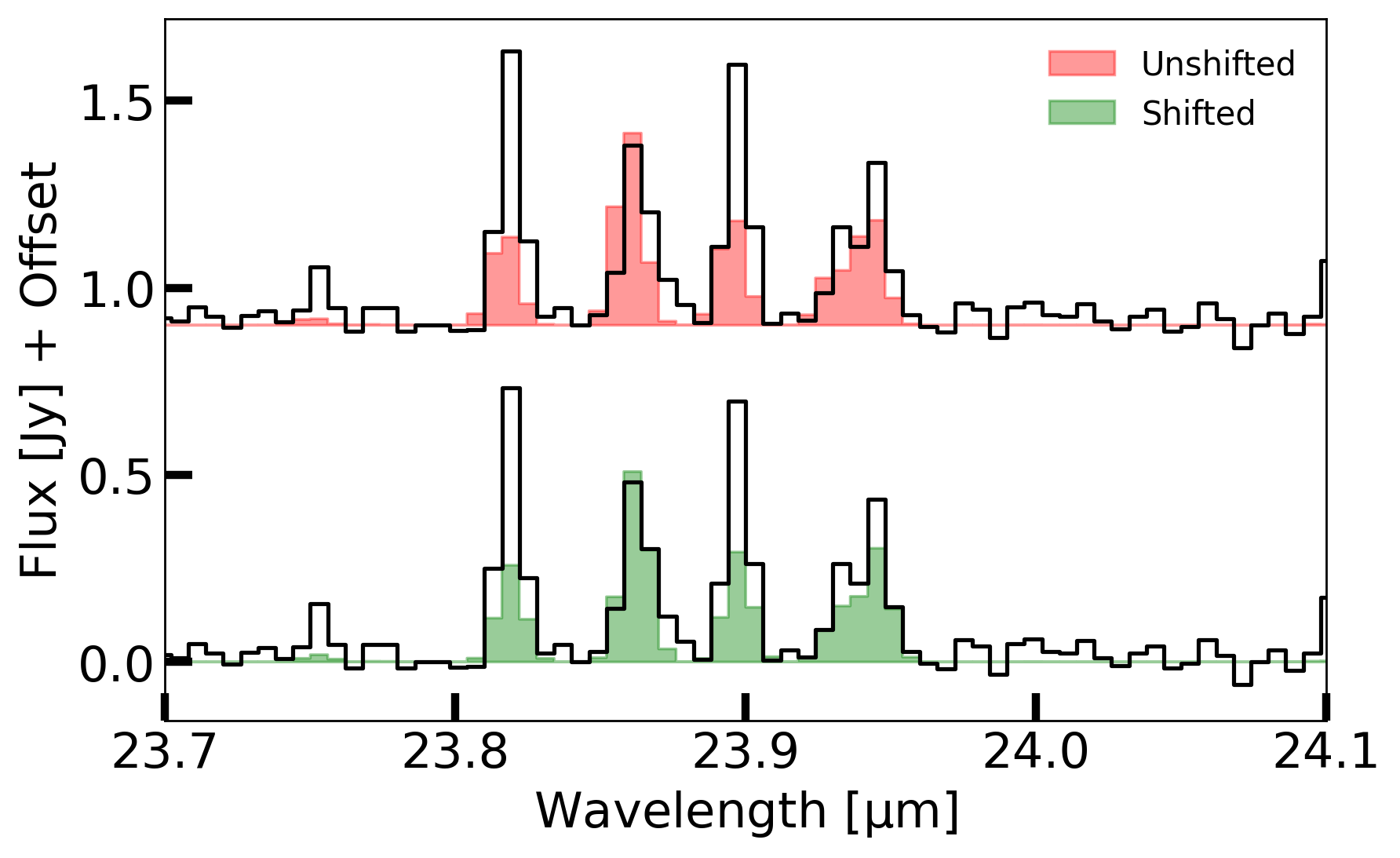}
    \caption{Comparison, using DR~Tau as a test case, between an unshifted slab model (red) and one shifted by the heliocentric velocity.}
    \label{fig:VHelShift}
\end{figure}

\section{Signal-to-noise ratios on the continuum and uncertainties for the observations}
\begin{table}[ht!]
    \centering
    \caption{S/N-ratios for each subband obtained from the \textit{Exposure Time Calculator}.}
    \begin{tabular}{c c c c c c c c c c c c c}
        \hline\hline
        Source & \multicolumn{12}{c}{ETC S/N-ratio in each subband} \\
               & 1A & 1B  & 1C & 2A & 2B & 2C & 3A & 3B & 3C & 4A & 4B & 4C \\
        \hline
        BP~Tau & 383.46 & 428.72 & 519.87 & 570.45 & 900.21 & 996.66 & 799.42 & 827.84 & 992.3 & 519.48 & 301.15 & 73.06 \\
        CX~Tau & 368.49 & 422.36 & 529.29 & 575.06 & 794.03 & 892.06 & 783.2 & 835.26 & 984.43 & 518.9 & 339.39 & 103.01 \\
        CY~Tau & 497.79 & 557.57 & 647.37 & 619.63 & 736.81 & 782.54 & 685.5 & 666.27 & 671.04 & 288.82 & 155.6 & 41.29 \\
        DN~Tau & 511.39 & 584.19 & 700.23 & 723.75 & 921.2 & 1046.72 & 961.64 & 1014.51 & 1143.0 & 622.85 & 427.47 & 137.28 \\
        DR~Tau & 558.15 & 624.65 & 755.79 & 767.27 & 994.22 & 1098.16 & 968.84 & 1006.86 & 1133.53 & 493.85 & 265.88 & 60.77 \\
        FT~Tau & 373.63 & 414.88 & 504.52 & 553.14 & 769.13 & 857.08 & 724.21 & 754.76 & 843.66 & 405.9 & 245.9 & 66.0 \\
        RNO~90 & 668.11 & 719.03 & 835.31 & 833.83 & 1075.98 & 1194.92 & 910.9 & 922.45 & 1057.8 & 435.4 & 219.53 & 45.73 \\
        XX~Cha & 383.02 & 423.56 & 506.03 & 538.19 & 740.69 & 822.55 & 720.9 & 753.22 & 862.37 & 457.42 & 283.29 & 78.24 \\
        \hline
    \end{tabular}
    \label{tab:S/N-Ratio}
\end{table}

\clearpage
\section{Slab model fit regions}
\begin{table}[ht!]
    \centering
    \caption{Fit regions used for the fitting of \ce{H_2O} transitions.}
    \begin{tabular}{c c c}
        \hline\hline
        Molecule & Sources & Fit regions$^\alpha$ \\
        \hline
        \ce{H_2O} & BP~Tau, CY~Tau, DR~Tau, FT~Tau, RNO~90, XX~Cha & 11.717-11.735, 12.56-12.57, 13.128-13.138, 13.2875-13.318, \\
        & & 13.495-13.513, 14.202-14.218, 14.34-14.355, 14.42-14.438,  \\
        & & 14.505-14.523, 14.68-14.695, 14.857-14.872, 14.888-14.905, \\
        & & 15.17-15.19, 15.41-15.425, 15.447-15.463, 15.567-15.583, \\
        & & 15.615-15.635, 15.717-15.732, 15.75-15.767, 15.958-15.975, \\
        & & 16.10-16.125, 16.263-16.28, 16.313-16.332, 16.535-16.553, \\
        & & 16.715-16.735, 16.975-16.993, 17.093-17.111, 17.135-17.15, \\
        & & 17.318-17.333, 17.348-17.366, 17.367-17.38, 17.397-17.415, \\
        & & 17.493-17.515, 17.558-17.575, 17.910-17.955, 18.240-18.270, \\
        & & 18.702-18.732, 19.230-19.266, 19.338-19.368, 19.572-19.608, \\
        & & 19.674-19.704, 19.925-19.95, 20.556-20.586 (5), \\
        & & 20.646-20.676 (15), 20.972-21.008 (5), 21.313-21.35 (5), \\
        & & 21.356-21.386 (5), 21.65-21.686 (5), 22.064-22.10 (5), \\
        & & 22.118-22.154 (5), 22.358-22.388 (5), 22.52-22.556 (5), \\
        & & 22.898-22.928 (5), 22.982-23.018 (5), 23.798-23.834 (15), \\
        & & 23.84-23.876 (5), 23.882-23.912 (15), 25.042-25.084 (10), \\
        & & 25.131-25.157 (5), 25.342-25.39 (5), 26.038-26.068 (5), \\
        & & 26.242-26.272 (5), 26.626-26.656 (5), 26.69-26.72 (10), \\
        & & 27.01-27.046 (15) \\
        & CX~Tau, DN~Tau & 11.717-11.735, 12.56-12.57, 13.128-13.138, 13.2875-13.318, \\
        & & 13.495-13.513, 15.958-15.975, 16.10-16.125, 16.263-16.28, \\
        & & 16.313-16.332, 16.535-16.553, 16.715-16.735, 16.975-16.993, \\
        & & 17.093-17.111, 17.135-17.15, 17.318-17.333, 17.348-17.366, \\
        & & 17.367-17.38, 17.397-17.415, 17.493-17.515, 17.558-17.575, \\
        & & 17.910-17.955, 18.240-18.270, 18.702-18.732, 19.230-19.266, \\
        & & 19.338-19.368, 19.572-19.608, 19.674-19.704, 19.925-19.95, \\
        & & 20.556-20.586 (5), 20.646-20.676 (15), 20.972-21.008 (5), \\
        & & 21.313-21.35 (5), 21.356-21.386 (5), 21.65-21.686 (5), \\
        & & 22.064-22.10 (5), 22.118-22.154 (5), 22.358-22.388 (5), \\
        & & 22.52-22.556 (5), 22.898-22.928 (5), 22.982-23.018 (5), \\
        & & 23.798-23.834 (15), 23.84-23.876 (5), 23.882-23.912 (15), \\
        & & 25.042-25.084 (10), 25.131-25.157 (5), 25.342-25.39 (5), \\
        & & 26.038-26.068 (5), 26.242-26.272 (5), 26.626-26.656 (5), \\
        & & 26.69-26.72 (10), 27.01-27.046 (15) \\
        \hline
    \end{tabular}
    \tablefoot{$^\alpha$: values listed in parenthesis in the fit regions denote the chosen weight, $w$, in Equation \ref{eq:Chi2}. If no value is listed, the weight is set to unity.}
    \label{tab:FitRegs}
\end{table}

\clearpage
\section{Fit results}
\begin{figure*}[ht!]
    \centering
    \includegraphics[width=0.95\textwidth]{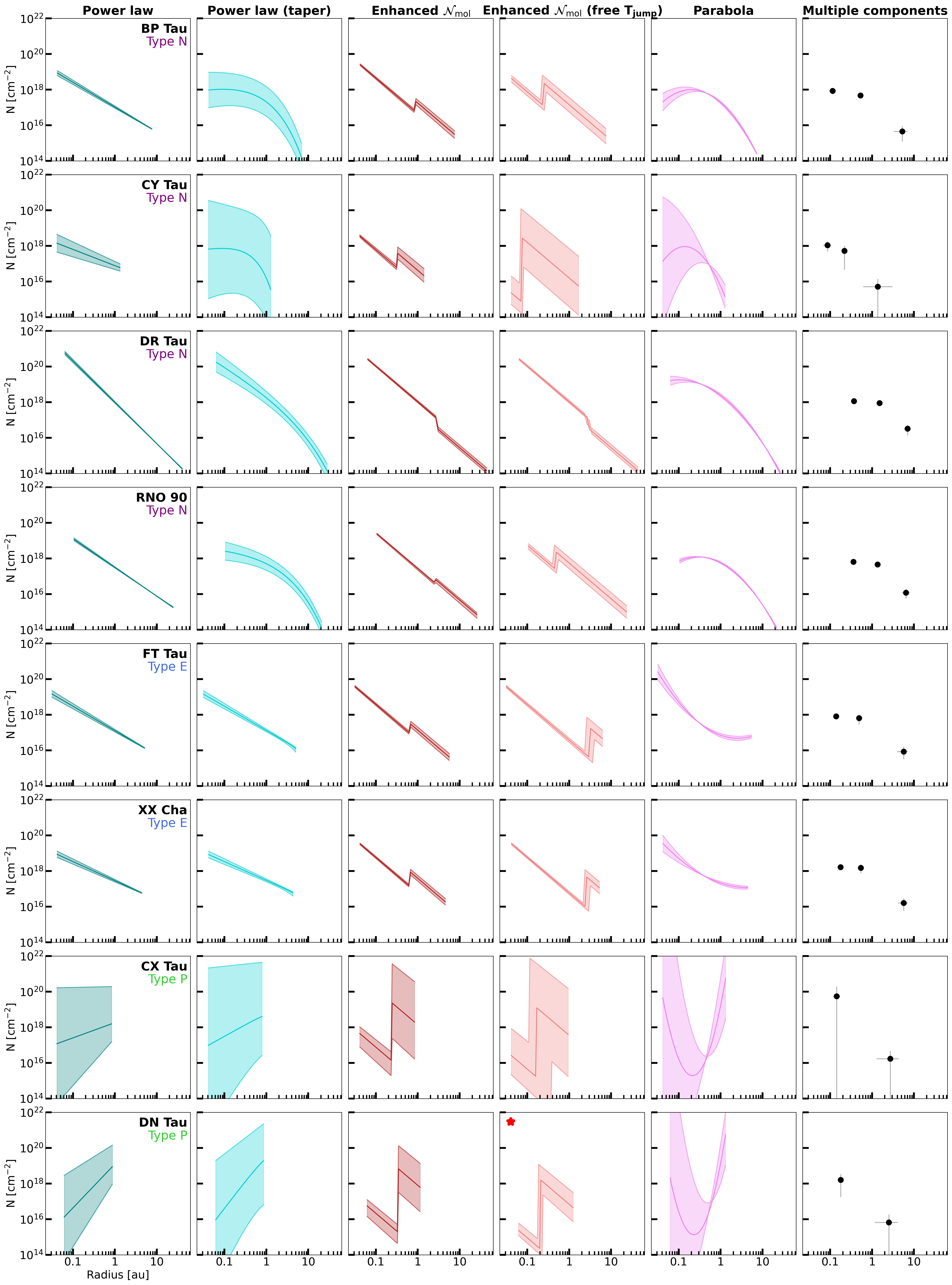}
    \caption{Similar to Figure \ref{fig:AllProfs-4.71}, but for the quadrature line widths. The shaded area is the 1$\sigma$-confidence interval, given the uncertainties listed in Tables \ref{tab:ProfileParams-Quad}.}
    \label{fig:AllProfs-Quad}
\end{figure*}

\begin{table*}[ht!]
    \centering
    \caption{Summary of the results: $\chi^2_\mathrm{red}$ values.}
    \begin{tabular}{c | c c c c c c c c c c | c c}
        \hline\hline
        Source & \multicolumn{2}{c}{I} & \multicolumn{2}{c}{II} & \multicolumn{2}{c}{III} & \multicolumn{2}{c}{IV} & \multicolumn{2}{c}{V} & \multicolumn{2}{c}{MC} \\
        & 4.71 & Quad. & 4.71 & Quad. & 4.71 & Quad. & 4.71 & Quad. & 4.71 & Quad. & 4.71 & Quad. \\
        \hline
        BP~Tau & 157.58 & 173.89 & 140.25 & 145.55 & 187.20 & 198.48 & \textbf{128.52} & 143.63 & 139.37 & 142.61 & 117.23 & 133.22 \\
        CX~Tau & 32.37 & 29.59 & 33.01 & 29.81 & 29.27 & 27.95 & 27.45 & 27.35 & \textbf{24.08} & 24.74 & 25.04 & 21.51 \\
        CY~Tau & 70.80 & 69.94 & 72.58 & 67.00 & 67.15 & 78.36 & \textbf{55.48} & 62.22 & 70.79 & 64.82 & 55.39 & 60.22 \\
        DN~Tau & 45.48 & 42.03 & 46.70 & 43.11 & 40.43 & 39.38 & 33.39 & \textbf{32.98} & 43.14 & 43.26 & 27.13 & 27.88 \\
        DR~Tau & 277.59 & 327.08 & 240.21 & 314.62 & 279.71 & 293.29 & 251.75 & 293.77 & \textbf{232.89} & 298.08 & 222.74 & 243.85 \\
        FT~Tau & 125.18 & 138.15 & 125.68 & 138.39 & 127.23 & 144.95 & \textbf{102.26} & 116.22 & 104.21 & 123.19 & 88.35 & 97.56 \\
        RNO~90 & 286.99 & 331.84 & \textbf{265.07} & 296.33 & 341.82 & 359.50 & 265.30 & 291.75 & 265.45 & 284.71 & 254.25 & 272.68 \\
        XX~Cha & 200.57 & 197.65 & 200.45 & 198.50 & 211.05 & 218.82 & 191.71 & 191.13 & \textbf{189.78} & 193.44 & 149.05 & 161.40 \\
        \hline
    \end{tabular}
    \label{tab:ResultsSum}
    \tablefoot{The Roman numbers have the following meanings: I - Power law fit, II - Power law with exponential taper, III - Jump abundance ($T_\mathrm{jump}$=400 K), IV - Jump abundance (free $T_\mathrm{jump}$), V - Parabola in log-space. MC stands for the multiple component analysis. \\
    The bold-faced entries highlight the fitted profile with the lowest value for $\chi^2_\mathrm{red}$.}
\end{table*}

\begin{table*}[ht!]
    \centering
    \caption{Fit values for the profiles with a line width of 4.71 km~s$^{-1}$.}
    \begin{tabular}{c | c c c c c c c c}
        \hline\hline
        & \multicolumn{3}{c|}{Power law} & \multicolumn{5}{c}{Power law with exponential taper} \\
        Source & $q$ & $\log_{10}\left(N_0\right)$ & $p$ & $q$ & $\log_{10}\left(N_0\right)$ & $p$ & $\log_{10}\left(R_\mathrm{c}\right)$ & $\phi$ \\
        \hline
        BP~Tau & 0.57$\pm$0.01 & 17.67$\pm$0.09 & 0.64$^{+0.08}_{-0.09}$ & 0.57$\pm$0.01 & 19.27$^{+0.81}_{-0.68}$ & -0.86$^{+0.52}_{-0.56}$ & -0.70$^{+0.39}_{-0.45}$ & 0.89$^{+0.43}_{-0.21}$ \\
        CX~Tau & 1.13$^{+0.10}_{-0.11}$ & 22.87$^{+2.55}_{-2.74}$ & -4.54$^{+2.26}_{-2.18}$ & 1.14$\pm$0.10 & 23.79$^{+2.49}_{-3.10}$ & -5.28$^{+2.57}_{-2.13}$ & -0.09$^{+0.74}_{-0.69}$ & 2.51$^{+1.66}_{-1.37}$ \\
        CY~Tau & 1.08$\pm$0.04 & 20.64$\pm$0.99 & -2.08$^{+0.81}_{-0.80}$ & 1.08$\pm$0.04 & 20.87$^{+1.13}_{-1.07}$ & -2.25$^{+0.86}_{-0.91}$ & -0.01$^{+0.69}_{-0.67}$ & 2.52$^{+1.65}_{-1.41}$ \\
        DN~Tau & 1.29$\pm$0.08 & 24.25$^{+1.47}_{-1.62}$ & -7.20$^{+1.63}_{-1.52}$ & 1.28$\pm$0.08 & 24.73$^{+1.45}_{-1.71}$ & -7.65$^{+1.72}_{-1.47}$ & -0.16$^{+0.78}_{-0.57}$ & 2.54$^{+1.60}_{-1.36}$ \\
        DR~Tau & 0.44$\pm$0.01 & 17.94$\pm$0.06 & 1.86$\pm$0.06 & 0.45$\pm$0.01 & 19.97$^{+0.26}_{-0.36}$ & -0.39$^{+0.30}_{-0.27}$ & -1.04$^{+0.24}_{-0.13}$ & 0.57$^{+0.06}_{-0.03}$ \\
        FT~Tau & 0.56$\pm$0.01 & 17.55$\pm$0.09 & 0.80$\pm$0.09 & 0.56$\pm$0.01 & 17.56$\pm$0.09 & 0.79$\pm$0.09 & 0.76$^{+0.17}_{-0.22}$ & 3.52$^{+1.03}_{-1.38}$ \\
        RNO~90 & 0.56$\pm$0.01 & 17.81$\pm$0.04 & 0.94$^{+0.06}_{-0.05}$ & 0.56$\pm$0.01 & 18.67$^{+0.56}_{-0.36}$ & -0.32$^{+0.38}_{-0.46}$ & -0.20$^{+0.30}_{-0.46}$ & 0.94$^{+0.38}_{-0.25}$ \\
        XX~Cha & 0.62$\pm$0.01 & 17.99$^{+0.10}_{-0.09}$ & 0.40$\pm$0.09 & 0.62$\pm$0.01 & 18.01$^{+0.09}_{-0.10}$ & 0.39$^{+0.09}_{-0.10}$ & 0.74$^{+0.18}_{-0.23}$ & 3.39$^{+1.12}_{-1.51}$ \\
        \hline\hline
        & \multicolumn{3}{c|}{Jump abundance ($T_\mathrm{jump}$=400 K)} & \multicolumn{4}{c}{Jump abundance ($T_\mathrm{jump}$ free)} & \\
        Source & $q$ & $\log_{10}\left(\mathcal{N}_\mathrm{mol}\right)$ & $\log_{10}\left(F_\mathrm{scale}\right)$ & $q$ & $T_\mathrm{jump}$ & $\log_{10}\left(\mathcal{N}_\mathrm{mol}\right)$ & $\log_{10}\left(F_\mathrm{scale}\right)$ & \\
        \hline
        BP~Tau & 0.58$\pm$0.01 & 42.42$\pm$0.03 & 1.44$^{+0.12}_{-0.11}$ & 0.53$\pm$0.01 & 843$\pm$28 & 38.71$^{+0.54}_{-0.38}$ & 4.42$^{+0.38}_{-0.54}$ & - \\
        CX~Tau & 1.02$\pm$0.07 & 39.70$^{+0.60}_{-1.00}$ & 5.02$^{+1.25}_{-1.36}$ & 0.99$^{+0.08}_{-0.07}$ & 599$^{+109}_{-127}$ & 38.60$^{+0.91}_{-0.72}$ & 4.32$^{+0.83}_{-0.84}$ & - \\
        CY~Tau & 1.05$\pm$0.03 & 41.33$\pm$0.06 & 3.02$^{+0.31}_{-0.30}$ & 0.87$^{+0.03}_{-0.02}$ & 989$^{+41}_{-42}$ & 38.25$^{+0.69}_{-0.58}$ & 3.88$^{+0.58}_{-0.68}$ & -\\
        DN~Tau & 1.04$\pm$0.05 & 39.53$^{+0.37}_{-0.60}$ & 4.45$^{+0.71}_{-0.61}$ & 1.03$^{+0.06}_{-0.05}$ & 593$^{+52}_{-53}$ & 38.55$^{+0.49}_{-0.37}$ & 4.08$^{+0.40}_{-0.49}$ & - \\
        DR~Tau & 0.42$\pm$0.01 & 43.98$^{+0.05}_{-0.02}$ & -0.13$\pm$0.08 & 0.45$\pm$0.01 & 854$^{+62}_{-98}$ & 43.13$^{+0.39}_{-0.26}$ & 1.16$^{+0.27}_{-0.38}$ & - \\
        FT~Tau & 0.54$\pm$0.01 & 42.27$\pm$0.04 & 1.06$\pm$0.08 & 0.53$\pm$0.01 & 247$\pm$18 & 42.30$\pm$0.04 & 1.84$^{+0.21}_{-0.18}$ & - \\
        RNO~90 & 0.55$\pm$0.01 & 43.25$\pm$0.02 & 0.98$\pm$0.06 & 0.55$\pm$0.01 & 655$^{+29}_{-28}$ & 42.83$^{+0.07}_{-0.08}$ & 1.09$\pm$0.08 & - \\
        XX~Cha & 0.61$\pm$0.01 & 42.33$\pm$0.04 & 1.45$\pm$0.09 & 0.59$\pm$0.01 & 258$^{+17}_{-16}$ & 42.39$\pm$0.03 & 2.19$^{+0.26}_{-0.20}$ & - \\
        \hline\hline
        & \multicolumn{4}{c|}{Parabola} & & & & \\
        Source & $q$ & $\log_{10}\left(N_0\right)$ & $\alpha$ & $\beta$ & & & & \\
        \hline
        BP~Tau & 0.57$\pm$0.01 & 17.40$^{+0.11}_{-0.12}$ & -2.56$^{+0.43}_{-0.49}$ & 0.79$^{+0.19}_{-0.17}$ & - & - & - & - \\
        CX~Tau & 0.77$^{+0.05}_{-0.04}$ & 23.80$^{+1.72}_{-1.97}$ & 21.93$^{+2.89}_{-4.05}$ & -5.80$^{+0.94}_{-0.72}$ & - & - & - & - \\
        CY~Tau & 1.04$^{+0.05}_{-0.04}$ & 15.07$^{+3.48}_{-2.21}$ & -8.65$^{+6.81}_{-4.55}$ & 2.22$^{+1.00}_{-1.43}$ & - & - & - & - \\
        DN~Tau & 1.25$\pm$0.08 & 16.15$^{+3.89}_{-2.61}$ & -11.01$^{+8.33}_{-5.91}$ & 4.41$^{+1.42}_{-1.97}$ & - & - & - & - \\
        DR~Tau & 0.45$\pm$0.01 & 18.28$\pm$0.06 & -2.21$\pm$0.08 & 0.79$\pm$0.08 & - & - & - & - \\
        FT~Tau & 0.55$\pm$0.01 & 17.65$\pm$0.11 & 0.97$^{+0.28}_{-0.27}$ & -0.69$\pm$0.10 & - & - & - & - \\
        RNO~90 & 0.56$\pm$0.01 & 17.97$^{+0.04}_{-0.05}$ & -1.34$^{+0.10}_{-0.11}$ & 0.56$^{+0.10}_{-0.09}$ & - & - & - & - \\
        XX~Cha & 0.61$\pm$0.01 & 18.08$^{+0.11}_{-0.10}$ & 1.03$^{+0.33}_{-0.31}$ & 0.59$\pm$0.12 & - & - & - & - \\
        \hline
    \end{tabular}
    \label{tab:ProfileParams-4.71}
\end{table*}

\begin{table*}[ht!] 
    \centering
    \caption{Fit values for the profiles with the quadrature line widths.}
    \begin{tabular}{c | c c c c c c c c}
        \hline\hline
        & \multicolumn{3}{c|}{Power law} & \multicolumn{5}{c}{Power law with exponential taper} \\
        Source & $q$ & $\log_{10}\left(N_0\right)$ & $p$ & $q$ & $\log_{10}\left(N_0\right)$ & $p$ & $\log_{10}\left(R_\mathrm{c}\right)$ & $\phi$ \\
        \hline
        BP~Tau & 0.44$\pm$0.01 & 17.02$\pm$0.06 & 1.39$^{+0.05}_{-0.06}$ & 0.44$\pm$0.01 & 19.10$^{+0.38}_{-0.49}$ & -0.60$^{+0.35}_{-0.29}$ & -1.08$^{+0.35}_{-0.22}$ & 0.57$^{+0.09}_{-0.04}$ \\
        CX~Tau & 0.75$^{+0.08}_{-0.06}$ & 18.26$^{+2.03}_{-0.87}$ & -0.85$^{+0.80}_{-1.81}$ & 0.77$^{+0.08}_{-0.06}$ & 18.80$^{+2.87}_{-1.18}$ & -1.31$^{+1.06}_{-2.61}$ & 0.26$^{+0.50}_{-0.62}$ & 2.42$^{+1.71}_{-1.42}$ \\
        CY~Tau & 0.65$\pm$0.02 & 16.89$^{+0.23}_{-0.22}$ & 0.91$^{+0.19}_{-0.20}$ & 0.66$\pm$0.02 & 18.39$^{+1.36}_{-1.14}$ & -0.37$^{+0.94}_{-1.04}$ & -0.63$^{+0.53}_{-0.49}$ & 1.12$^{+0.93}_{-0.33}$ \\
        DN~Tau & 0.85$\pm$0.05 & 19.10$^{+1.15}_{-0.91}$ & -2.46$^{+0.98}_{-1.28}$ & 0.86$\pm$0.05 & 19.67$^{+1.81}_{-1.15}$ & -3.06$^{+1.25}_{-1.97}$ & 0.11$^{+0.61}_{-0.69}$ & 2.09$^{+1.87}_{-1.11}$ \\
        DR~Tau & 0.35$\pm$0.01 & 17.99$^{+0.05}_{-0.04}$ & 2.32$\pm$0.04 & 0.37$\pm$0.01 & 18.67$^{+0.21}_{-0.19}$ & 1.40$\pm$0.26 & 0.02$^{+0.31}_{0.26}$ & 0.53$^{+0.05}_{-0.02}$ \\
        FT~Tau & 0.44$\pm$0.01 & 17.11$\pm$0.07 & 1.37$\pm$0.07 & 0.45$\pm$0.01 & 17.12$^{+0.08}_{-0.07}$ & 1.36$\pm$0.08 & 0.90$^{+0.07}_{-0.10}$ & 3.84$^{+0.82}_{-1.24}$ \\
        RNO~90 & 0.41$\pm$0.01 & 17.50$^{+0.03}_{-0.04}$ & 1.61$\pm$0.04 & 0.43$\pm$0.01 & 18.58$^{+0.21}_{-0.22}$ & 0.09$\pm$0.23 & -0.57$^{+0.24}_{-0.16}$ & 0.53$^{+0.05}_{-0.02}$ \\
        XX~Cha & 0.49$\pm$0.01 & 17.46$\pm$0.07 & 1.07$^{+0.08}_{-0.07}$ & 0.49$\pm$0.01 & 17.48$^{+0.08}_{-0.07}$ & 1.05$\pm$0.08 & 0.90$^{+0.07}_{-0.11}$ & 3.72$^{+0.91}_{-1.53}$ \\
        \hline\hline
        & \multicolumn{3}{c|}{Jump abundance ($T_\mathrm{jump}$=400 K)} & \multicolumn{4}{c}{Jump abundance ($T_\mathrm{jump}$ free)} & \\
        Source & $q$ & $\log_{10}\left(\mathcal{N}_\mathrm{mol}\right)$ & $\log_{10}\left(F_\mathrm{scale}\right)$ & $q$ & $T_\mathrm{jump}$ & $\log_{10}\left(\mathcal{N}_\mathrm{mol}\right)$ & $\log_{10}\left(F_\mathrm{scale}\right)$ & \\
        \hline
        BP~Tau & 0.44$\pm$0.01 & 42.78$\pm$0.04 & 0.60$\pm$0.08 & 0.44$\pm$0.01 & 694$^{+39}_{-36}$ & 42.01$^{+0.15}_{-0.20}$ & 1.28$^{+0.20}_{-0.16}$ & - \\
        CX~Tau & 0.75$\pm$0.05 & 40.58$^{+0.34}_{-0.72}$ & 3.26$^{+1.73}_{-1.11}$ & 0.72$\pm$0.05 & 549$^{+165}_{-255}$ & 39.36$^{+1.48}_{-1.04}$ & 3.89$^{+0.89}_{-1.07}$ & - \\
        CY~Tau & 0.64$\pm$0.02 & 41.72$\pm$0.05 & 0.83$^{+0.22}_{-0.19}$ & 0.61$\pm$0.02 & 1044$^{+53}_{-61}$ & 38.55$^{+0.93}_{-0.63}$ & 3.60$^{+0.63}_{-0.91}$ & - \\
        DN~Tau & 0.78$\pm$0.03 & 40.16$^{+0.33}_{-0.57}$ & 3.57$^{+0.89}_{-0.65}$ & 0.75$\pm$0.03 & 602$^{+57}_{-56}$ & 38.80$^{+0.39}_{-0.27}$ & 3.88$^{+0.33}_{-0.40}$ & - \\
        DR~Tau & 0.35$\pm$0.01 & 44.29$^{+0.04}_{-0.03}$ & -0.57$\pm$0.06 & 0.35$\pm$0.01 & 393$^{+19}_{-18}$ & 44.29$\pm$0.04 & -0.57$\pm$0.07 & - \\
        FT~Tau & 0.44$\pm$0.01 & 42.66$\pm$0.05 & 0.55$\pm$0.07 & 0.43$\pm$0.01 & 205$^{+20}_{-17}$ & 42.66$\pm$0.06 & 1.67$^{+0.32}_{-0.28}$ & - \\
        RNO~90 & 0.41$\pm$0.01 & 43.54$^{+0.01}_{-0.05}$ & 0.23$^{+0.06}_{-0.05}$ & 0.42$\pm$0.01 & 779$^{+42}_{-40}$ & 42.83$^{+0.13}_{-0.17}$ & 1.00$^{+0.16}_{-0.14}$ & - \\
        XX~Cha & 0.48$\pm$0.01 & 42.77$\pm$0.04 & 0.83$\pm$0.08 & 0.47$\pm$0.01 & 214$^{+16}_{-14}$ & 42.79$^{+0.05}_{-0.04}$ & 1.73$^{+0.24}_{-0.21}$ & - \\
        \hline\hline
        & \multicolumn{4}{c|}{Parabola} & & & & \\
        Source & $q$ & $\log_{10}\left(N_0\right)$ & $\alpha$ & $\beta$ & & & & \\
        \hline
        BP~Tau & 0.44$\pm$0.01 & 17.28$\pm$0.06 & -2.06$^{+0.13}_{-0.12}$ & 0.64$\pm$0.09 & - & - & - & - \\
        CX~Tau & 0.65$^{+0.05}_{-0.04}$ & 19.22$^{+1.57}_{-1.72}$ & 11.71$^{+3.18}_{-4.74}$ & -3.78$^{+1.30}_{-0.90}$ & - & - & - & - \\
        CY~Tau & 0.66$\pm$0.02 & 15.75$^{+0.50}_{-0.49}$ & -5.19$^{+1.51}_{-1.56}$ & 1.32$^{+0.47}_{-0.44}$ & - & - & - & - \\
        DN~Tau & 0.74$^{+0.07}_{-0.03}$ & 19.14$^{+1.08}_{-1.13}$ & 12.57$^{+2.29}_{-5.42}$ & -4.28$^{+2.09}_{-0.78}$ & - & - & - & - \\
        DR~Tau & 0.37$\pm$0.01 & 18.38$^{+0.08}_{-0.07}$ & -1.77$\pm$0.09 & 0.39$^{+0.07}_{-0.06}$ & - & - & - & - \\
        FT~Tau & 0.44$\pm$0.01 & 16.85$^{+0.10}_{-0.11}$ & -0.84$\pm$0.11 & -0.44$\pm$0.07 & - & - & - & - \\
        RNO~90 & 0.42$\pm$0.01 & 17.77$^{+0.04}_{-0.05}$ & -1.22$\pm$0.06 & 0.50$^{+0.04}_{-0.05}$ & - & - & - & - \\
        XX~Cha & 0.48$\pm$0.01 & 17.33$\pm$0.08 & -0.79$\pm$0.12 & -0.26$\pm$0.09 & - & - & - & - \\
        \hline
    \end{tabular}
    \label{tab:ProfileParams-Quad}
\end{table*}

\begin{table*}
    \centering
    \caption{Fit values for the multiple component fits.}
    \begin{tabular}{c | c c c c c c c c c}
        \hline\hline
        \multicolumn{9}{c}{Fixed 4.71 line width} \\
         & BP~Tau & CX~Tau & CY~Tau & DN~Tau & DR~Tau & FT~Tau & RNO~90 & XX~Cha \\
        \hline
        $T_1$                       & 933$^{+136}_{-101}$ & 425$^{+114}_{-77}$ & 821$^{+150}_{-86}$    & 486$^{+53}_{-52}$    & 906$^{+64}_{-55}$ & 920$^{+56}_{-47}$ & 953$^{+66}_{-54}$ & 765$^{+43}_{-35}$ \\
        $\log_{10}\left(N_1\right)$ & 18.5$\pm$0.2        & 20.3$\pm$1.1       & 18.7$\pm$0.2          & 18.7$^{+0.5}_{-0.4}$ & 18.6$\pm$0.1      & 18.5$\pm$0.1      & 18.4$\pm$0.1      & 18.8$\pm$0.1 \\
        $R_1$                       & 0.12$\pm$0.02       & 0.13$\pm$0.03      & 0.08$^{+0.02}_{-0.1}$ & 0.16$\pm$0.02        & 0.39$\pm$0.04     & 0.13$\pm$0.01     & 0.34$\pm$0.03     & 0.18$^{+0.01}_{-0.02}$ \\
        \hline
        $T_2$                       & 516$^{+15}_{-17}$ & 196$^{+42}_{-31}$     & 535$^{+43}_{-56}$       & 239$^{+30}_{-31}$      & 487$^{+15}_{-16}$ & 364$^{+49}_{-33}$    & 513$\pm$13    & 357$^{+55}_{-38}$ \\
        $\log_{10}\left(N_2\right)$ & 18.2$\pm$0.1      & 16.3$^{+1.1}_{-0.7}$   & 18.1$\pm$0.4           & 15.8$^{+1.1}_{-0.6}$   & 18.5$\pm$0.1      & 18.3$^{+0.3}_{-0.2}$ & 18.2$\pm$0.1  & 18.7$\pm$0.2 \\
        $R_2$                       & 0.46$\pm$0.02     & 4.60$^{+3.54}_{-3.19}$ & 0.17$^{+0.03}_{-0.02}$ & 4.21$^{+3.37}_{-2.94}$ & 1.24$\pm$0.04     & 0.44$\pm$0.08        & 1.12$\pm$0.05 & 0.5$^{+0.08}_{-0.07}$ \\
        \hline
        $T_3$                       & 259$^{+18}_{-19}$      & - & 268$^{+67}_{-79}$      & - & 247$^{+22}_{-23}$      & 217$^{+14}_{-15}$      & 295$\pm$18             & 207$\pm$12 \\
        $\log_{10}\left(N_3\right)$ & 15.7$^{+0.8}_{-0.4}$   & - & 15.6$^{+0.8}_{-0.5}$   & - & 16.7$^{+0.5}_{-0.4}$   & 16.0$^{+0.6}_{-0.4}$   & 16.4$\pm$0.5           & 16.3$^{+0.6}_{-0.4}$ \\
        $R_3$                       & 7.59$^{+4.98}_{-4.57}$ & - & 2.45$^{+3.13}_{-1.55}$ & - & 8.39$^{+4.41}_{-3.65}$ & 9.00$^{+4.08}_{-4.28}$ & 6.97$^{+5.22}_{-3.03}$ & 8.79$^{4.21}_{-4.19}$ \\
        \hline
        \hline
        \multicolumn{9}{c}{Quadrature line widths} \\
         & BP~Tau & CX~Tau & CY~Tau & DN~Tau & DR~Tau & FT~Tau & RNO~90 & XX~Cha \\
        \hline
        $T_1$                       & 959$^{+150}_{-107}$    & 415$^{+113}_{-75}$ & 836$^{+169}_{-100}$  & 481$^{+56}_{-52}$    & 921$^{+57}_{-43}$      & 903$^{+60}_{-46}$ & 940$^{+59}_{-48}$ & 767$^{+49}_{-38}$ \\
        $\log_{10}\left(N_1\right)$ & 17.9$\pm$0.2           & 19.7$\pm$1.1       & 18.0$^{+0.3}_{-0.2}$ & 18.2$^{+0.5}_{-0.4}$ & 18.1$\pm$0.1           & 17.9$\pm$0.1      & 17.8$\pm$0.1      & 18.2$\pm$0.1 \\
        $R_1$                       & 0.19$^{+0.05}_{-0.04}$ & 0.25$\pm$0.06      & 0.13$\pm$0.04        & 0.30$\pm$0.04        & 0.68$^{+0.05}_{-0.06}$ & 0.25$\pm$0.02     & 0.62$\pm$0.06     & 0.31$^{+0.03}_{-0.04}$ \\
        \hline
        $T_2$                       & 517$^{+16}_{-17}$ & 192$^{+41}_{-28}$      & 530$^{+50}_{-65}$      & 231$^{+31}_{-32}$      & 491$\pm$15    & 382$^{+60}_{-48}$      & 511$^{+13}_{-14}$ & 389$^{+49}_{-54}$ \\
        $\log_{10}\left(N_2\right)$ & 17.7$\pm$0.1      & 16.2$^{+0.8}_{-0.6}$   & 17.7$\pm$0.4           & 15.8$^{+0.8}_{-0.5}$   & 18.0$\pm$0.1  & 17.8$^{+0.3}_{-0.2}$   & 17.7$\pm$0.1      & 18.2$\pm$0.2 \\
        $R_2$                       & 0.87$\pm$0.03     & 5.19$^{+3.15}_{-2.87}$ & 0.30$^{+0.04}_{-0.03}$ & 4.75$^{+3.20}_{-2.72}$ & 2.34$\pm$0.08 & 0.74$^{+0.17}_{-0.13}$ & 2.10$\pm$0.08     & 0.76$^{+0.15}_{-0.09}$ \\
        \hline
        $T_3$                       & 250$^{+17}_{-18}$      & - & 278$^{+66}_{-82}$      & - & 242$^{+19}_{-20}$      & 215$^{+13}_{-14}$       & 289$^{+16}_{-17}$       & 202$\pm$11 \\
        $\log_{10}\left(N_3\right)$ & 15.7$^{+0.5}_{-0.3}$   & - & 15.7$^{+0.8}_{-0.5}$   & - & 16.5$\pm$0.3           & 15.9$^{+0.4}_{-0.3}$    & 16.1$^{+0.3}_{-0.2}$    & 16.2$^{+0.4}_{-0.3}$ \\
        $R_3$                       & 9.67$^{+3.64}_{-3.93}$ & - & 2.42$^{+3.40}_{-1.50}$ & - & 11.8$^{+2.21}_{-2.62}$ & 10.62$^{+3.01}_{-3.48}$ & 10.84$^{+2.80}_{-2.76}$ & 10.60$^{+2.98}_{-3.03}$ \\
        \hline
    \end{tabular}
    \label{tab:MC-Results}
\end{table*}

\clearpage
\begin{figure}[ht!]
    \centering
    \includegraphics[width=0.89\textwidth]{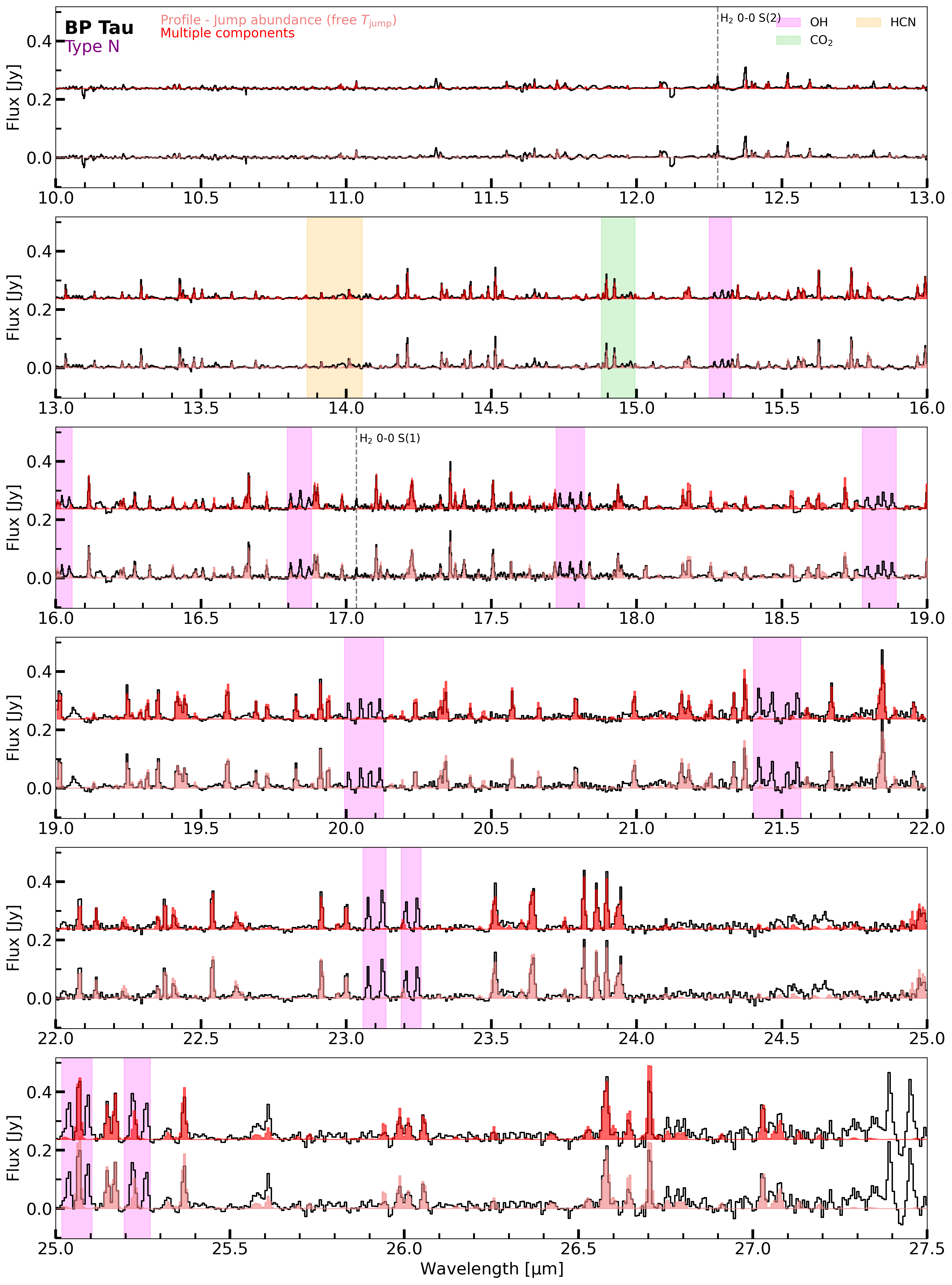}
    \caption{Comparison between the full models of the best-fitting parametric model (bottom fit) and of the multiple components (red, top fit) for BP~Tau. The colour of the best-fitting profile matches that of Figure \ref{fig:AllProfs-4.71}. Indicated are also the molecular features from \ce{OH} (magenta), \ce{CO_2} (green), and \ce{HCN} (orange). The vertical lines indicate the S(1) and S(2) transitions of \ce{H_2}.}
    \label{fig:BPTau-ProfFit}
\end{figure}
\begin{figure}[ht!]
    \centering
    \includegraphics[width=0.9\textwidth]{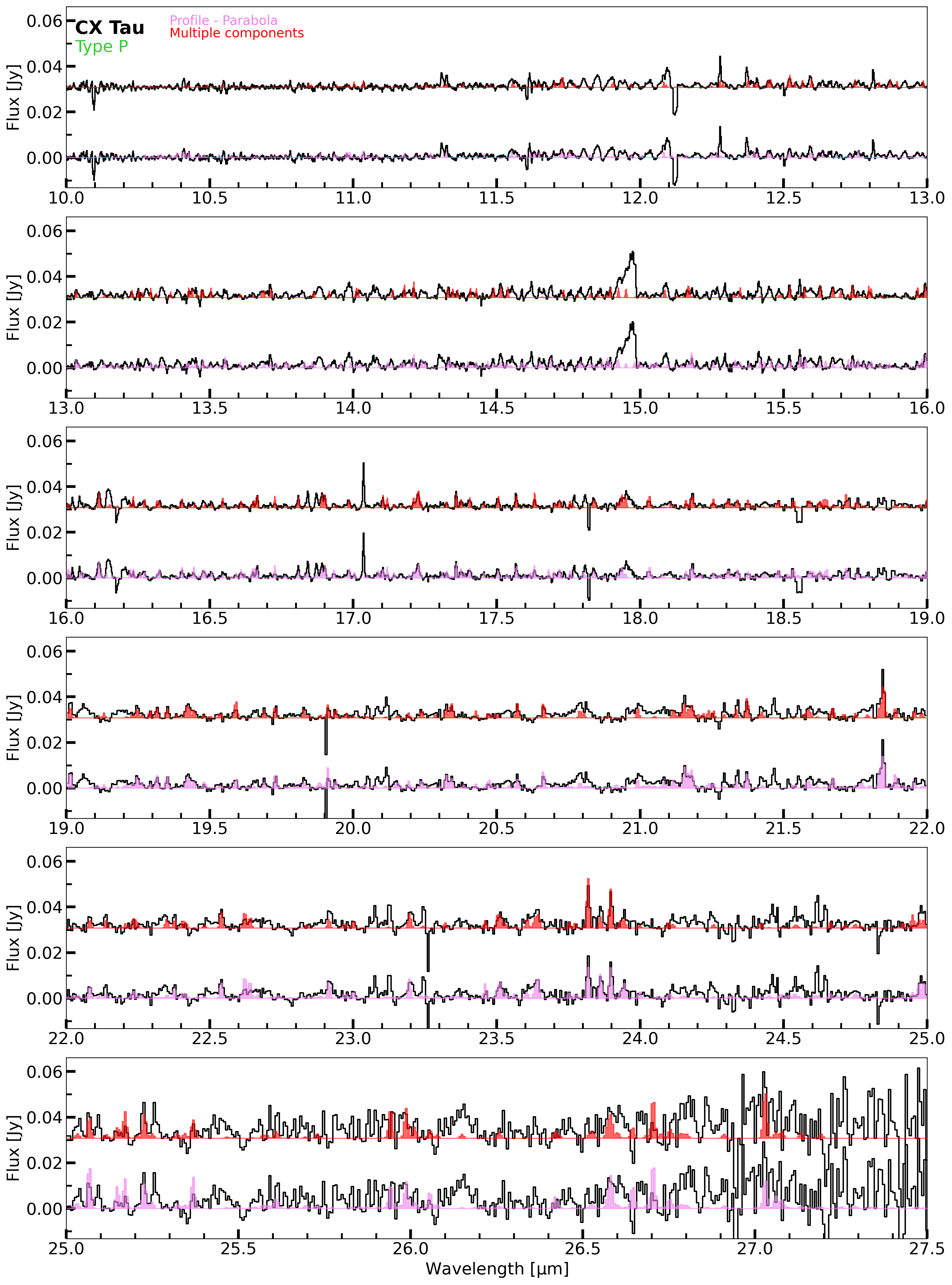}
    \caption{Similar as Figure \ref{fig:BPTau-ProfFit}, but for CX~Tau. We refer the reader to \citet{VlasblomEA24b} for the other molecular emission features.}
    \label{fig:CXTau-ProfFit}
\end{figure}
\begin{figure}[ht!]
    \centering
    \includegraphics[width=0.9\textwidth]{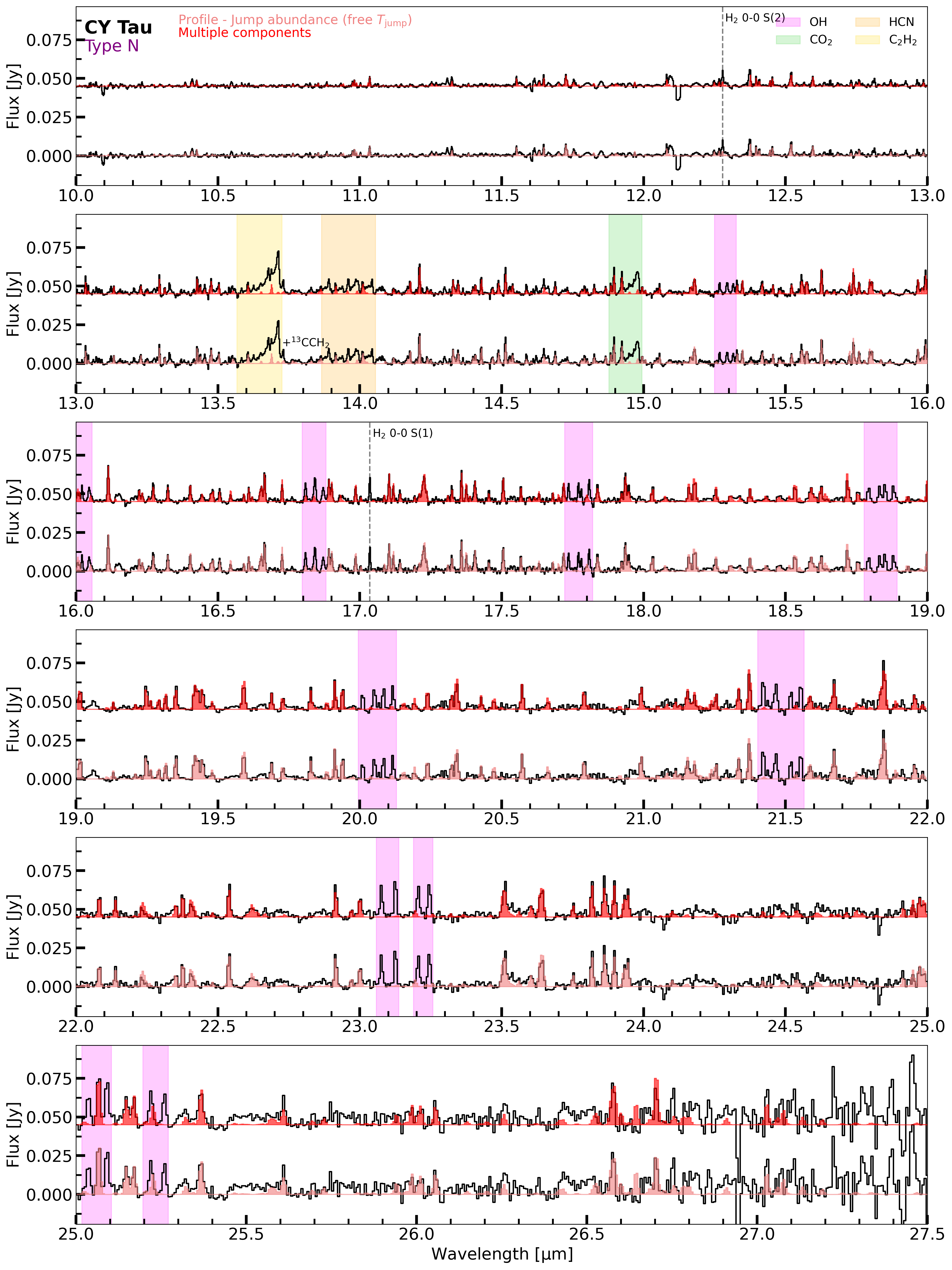}
    \caption{Similar as Figure \ref{fig:BPTau-ProfFit}, but for CY~Tau. Highlighted in yellow is the \ce{C_2H_2} emission feature, while the approximate location of \ce{^{13}CCH_2} is also shown.}
    \label{fig:CYTau-ProfFit}
\end{figure}
\begin{figure}[ht!]
    \centering
    \includegraphics[width=0.9\textwidth]{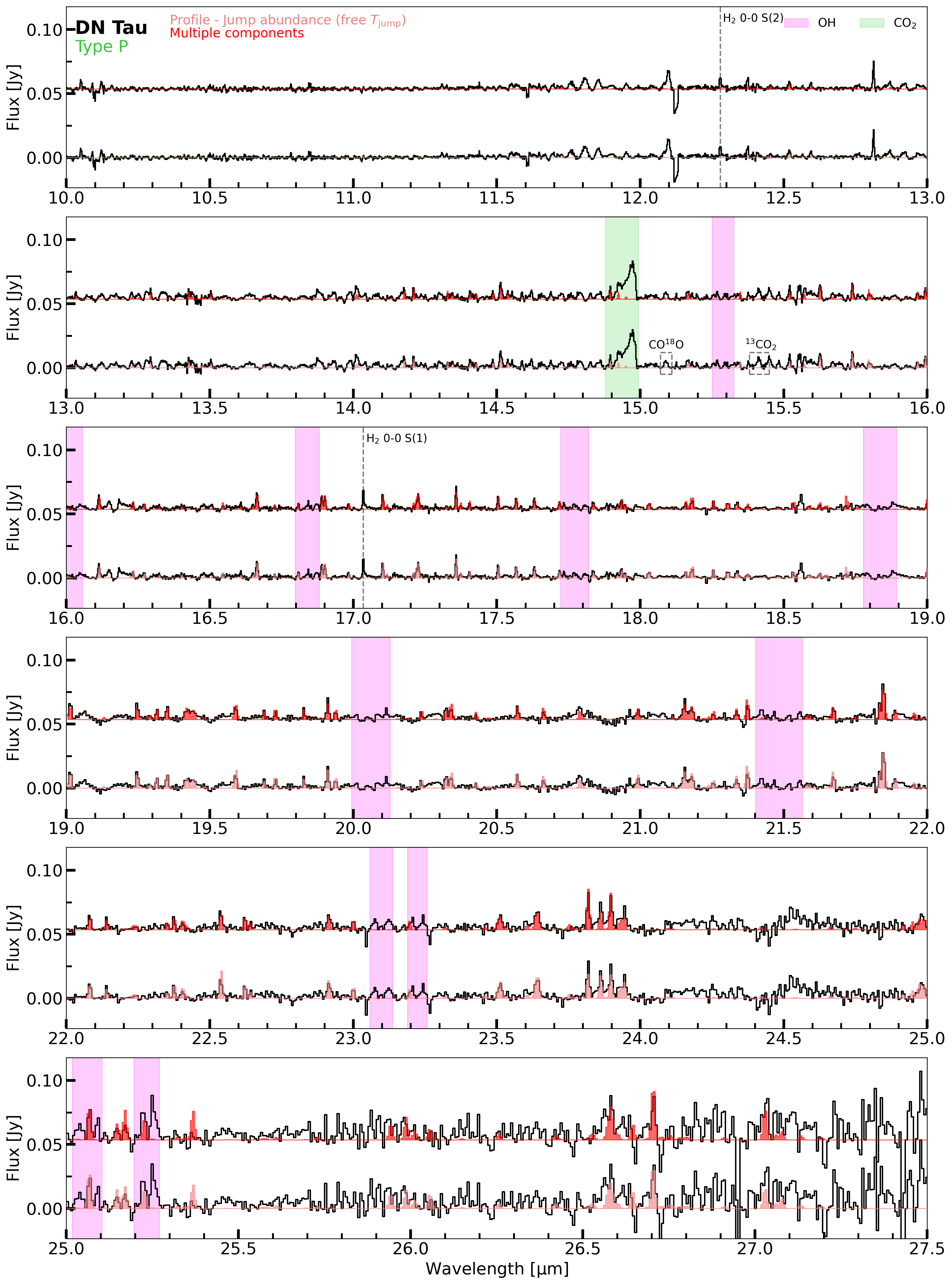}
    \caption{Similar as Figure \ref{fig:BPTau-ProfFit}, but for DN~Tau. The boxes indicate the detection of \ce{^{13}CO_2} and the potential detection of \ce{CO^{18}O}.}
    \label{fig:DNTau-ProfFit}
\end{figure}
\begin{figure}[ht!]
    \centering
    \includegraphics[width=0.9\textwidth]{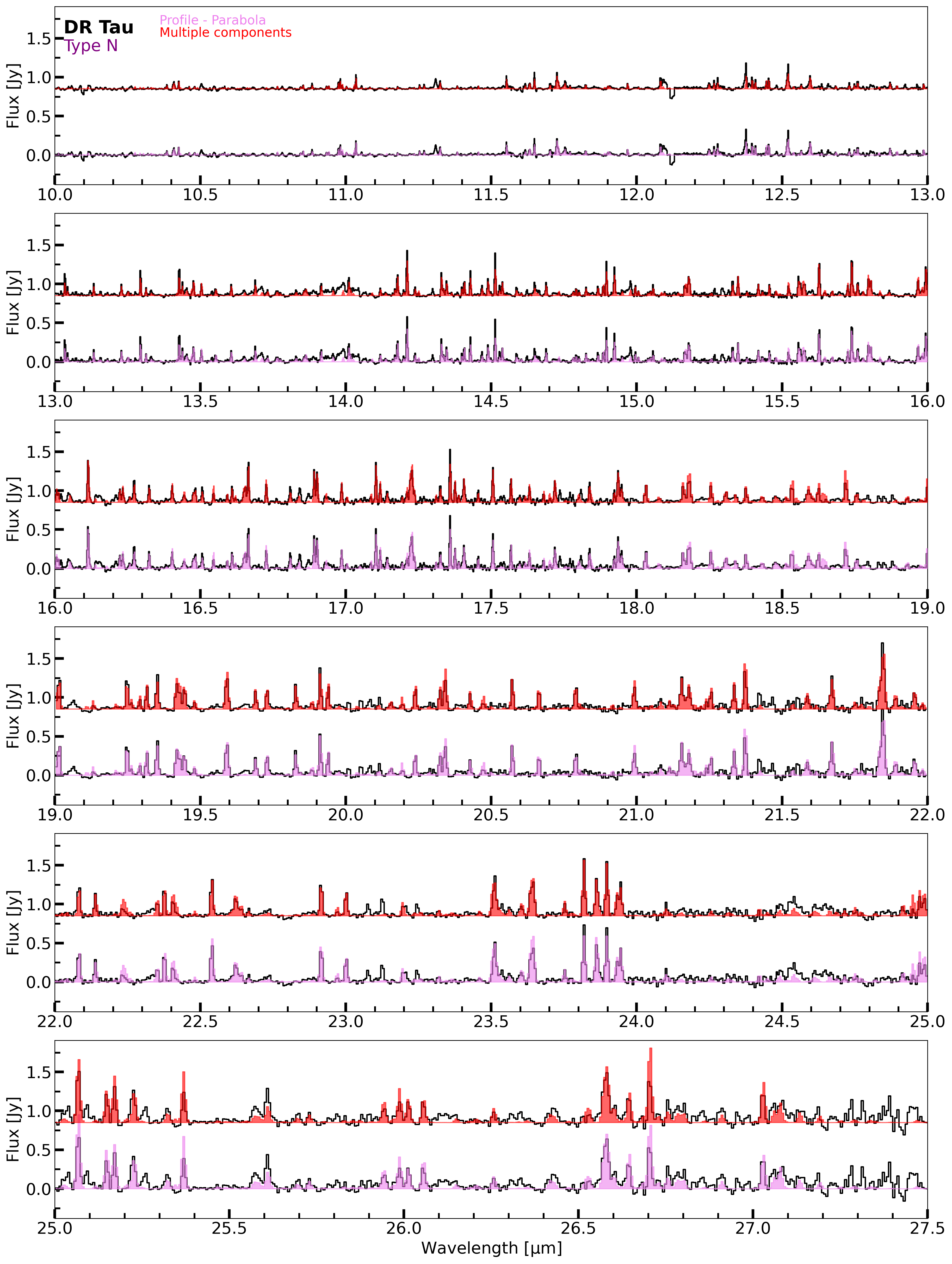}
    \caption{Similar as Figure \ref{fig:BPTau-ProfFit}, but for DR~Tau. We refer the reader to \citet{TemminkEA24} and \citet{TemminkEA24b} for the other molecular emission features.}
    \label{fig:DRTau-ProfFit}
\end{figure}
\begin{figure}[ht!]
    \centering
    \includegraphics[width=0.9\textwidth]{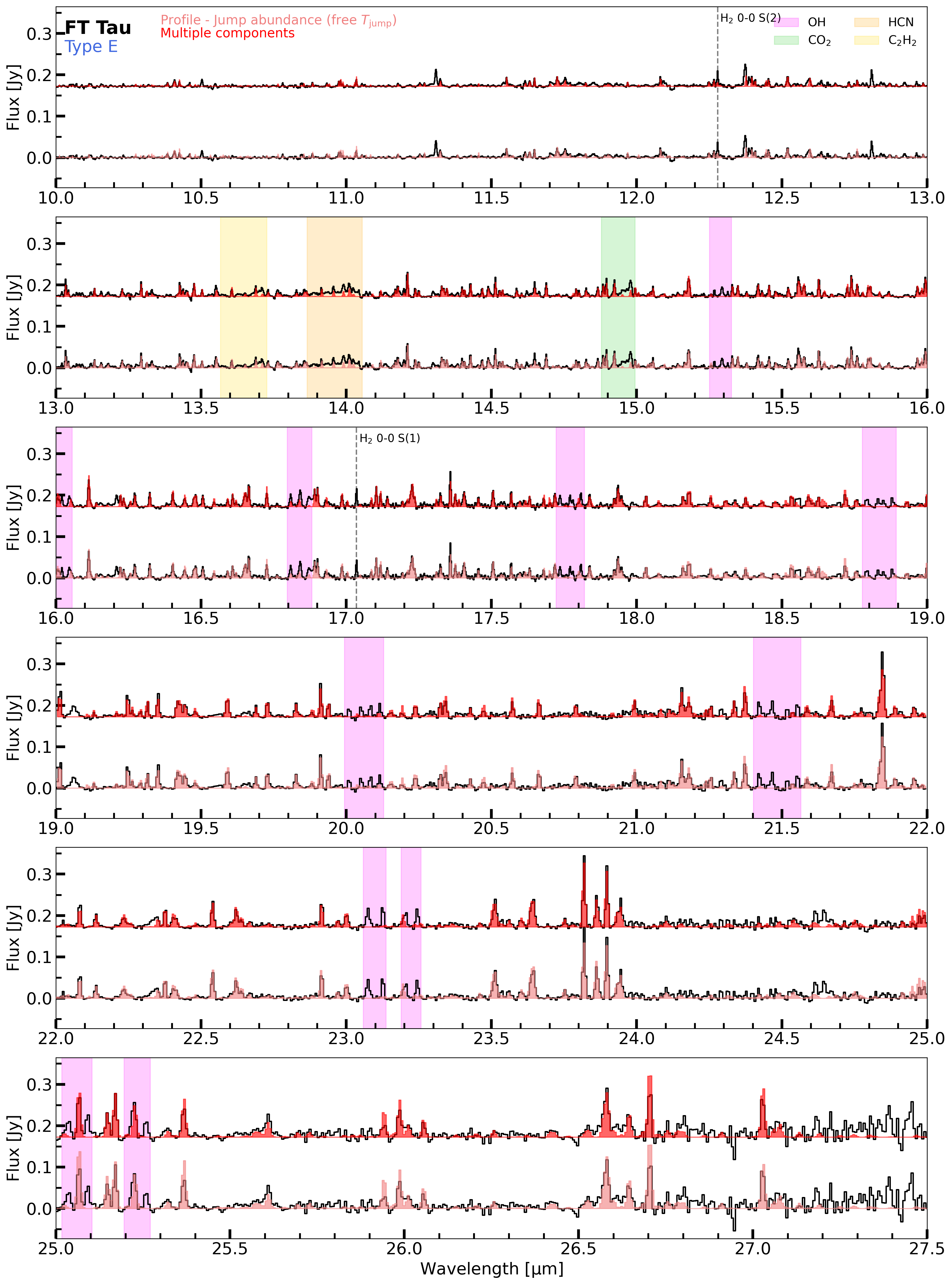}
    \caption{Similar as Figure \ref{fig:BPTau-ProfFit}, but for FT~Tau. Highlighted in yellow is the \ce{C_2H_2} emission feature.}
    \label{fig:FTTau-ProfFit}
\end{figure}
\begin{figure}[ht!]
    \centering
    \includegraphics[width=0.9\textwidth]{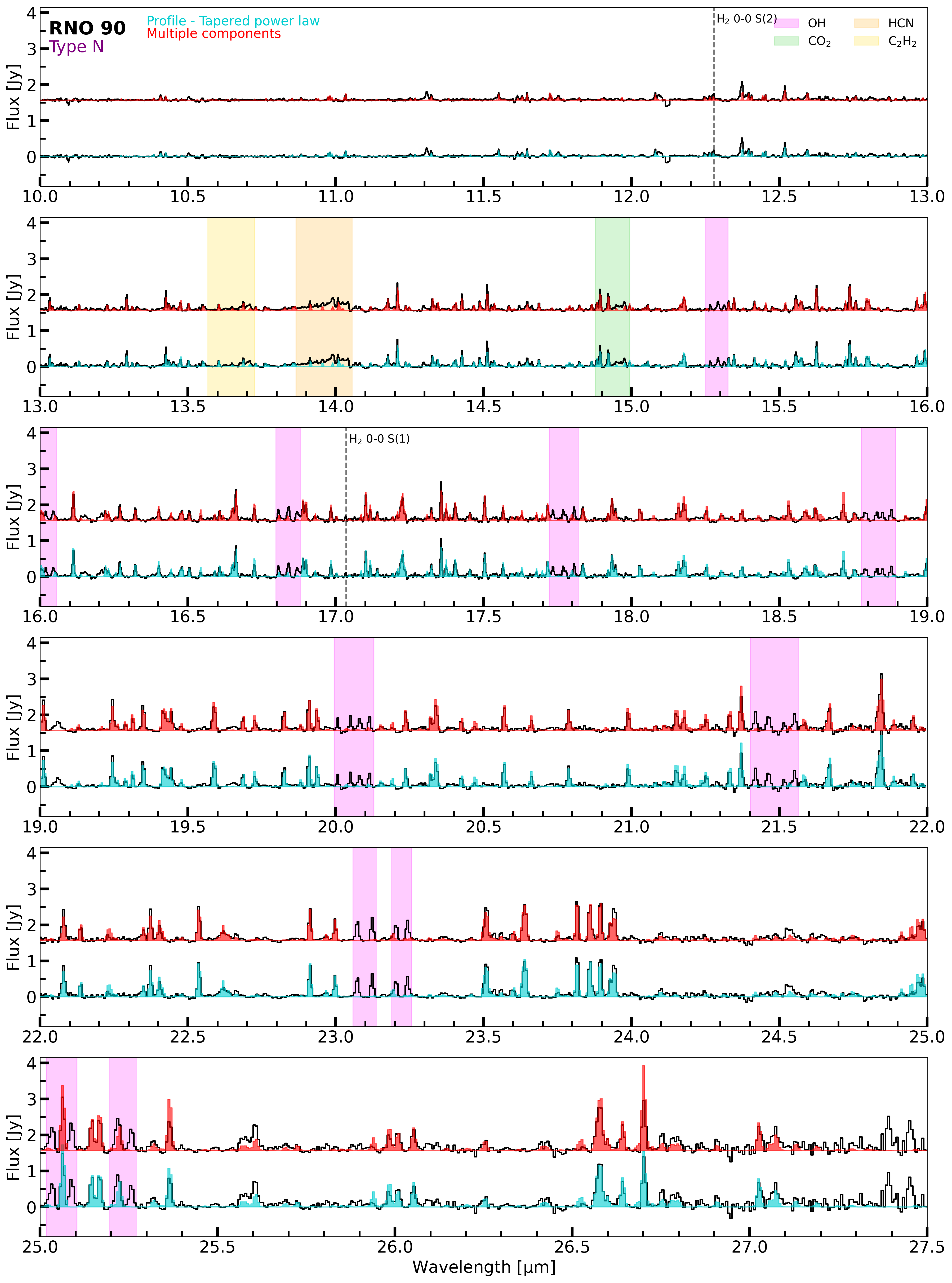}
    \caption{Similar as Figure \ref{fig:BPTau-ProfFit}, but for RNO~90. Highlighted in yellow is the \ce{C_2H_2} emission feature.}
    \label{fig:RNO90-ProfFit}
\end{figure}
\begin{figure}[ht!]
    \centering
    \includegraphics[width=0.9\textwidth]{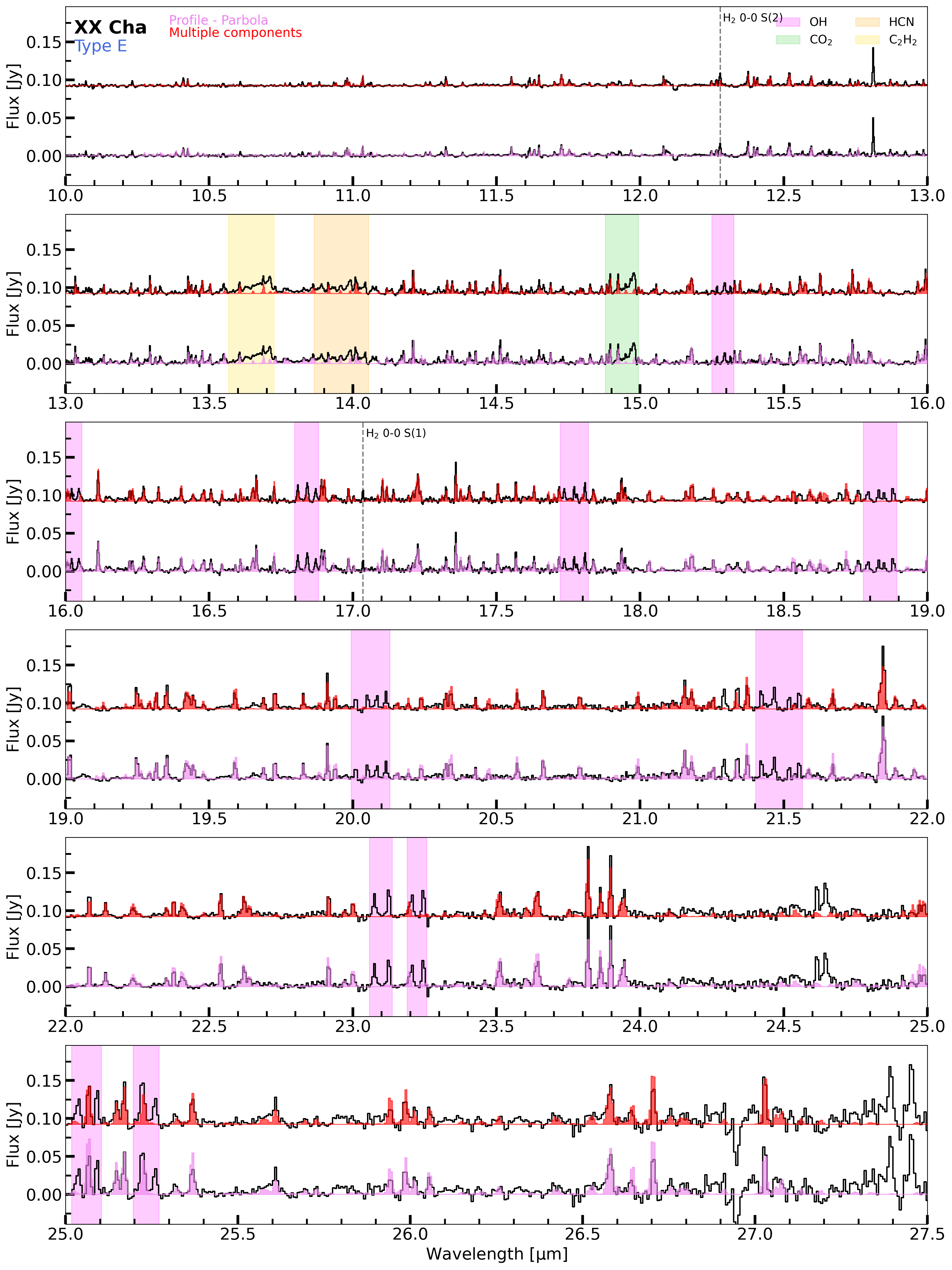}
    \caption{Similar as Figure \ref{fig:BPTau-ProfFit}, but for XX~Cha. Highlighted in yellow is the \ce{C_2H_2} emission feature.}
    \label{fig:XXCha-ProfFit}
\end{figure}

\clearpage
\section{Other molecular species} \label{sec:OMSpecies}
In this following section, we discuss the observed molecular species, aside from \ce{H_2O}, in each disk (Section \ref{sec:MolSource}) and we discuss the tentative detection of \ce{CH_4} in CY~Tau (Sectoin \ref{sec:CH4}). These features are also highlighted in Figures \ref{fig:BPTau-ProfFit}-\ref{fig:XXCha-ProfFit}.

\subsection{Molecules per source} \label{sec:MolSource}
\indent \textit{BP~Tau}: Aside from the pure rotational \ce{H_2O} lines, the spectrum of BP~Tau also contains strong emission of the ro-vibrational transitions and \ce{CO} emission at the shortest wavelengths (4.9-5.3 $\mathrm{\mu}$m). Other molecular species, such as \ce{CO_2} and \ce{HCN}, are only weakly detected, while \ce{OH} transitions are strongly detected above $>$15 $\mathrm{\mu}$m. Both the \ce{H_2} 0-0 S(1) (at 17.30484 $\mathrm{\mu}$m) and S(2) (at 12.27861 $\mathrm{\mu}$m) are detected. The detected molecular species at $>$10 $\mathrm{\mu}$m are also highlighted in Figure \ref{fig:BPTau-ProfFit}. \\
\indent \textit{CY~Tau}: The strongest molecular feature in the spectrum of CY~Tau belongs to \ce{C_2H_2}, also visible in Figure \ref{fig:Sample}. Its isotopologue, \ce{^{13}CCH_2} is also well detected. Other molecular species include \ce{CO}, the ro-vibrational lines of \ce{H_2O}, \ce{OH}, \ce{HCN}, and \ce{CO_2}. Both the \ce{H_2} S(1) and S(2) transitions are also detected. Additionally, we present a potential detection of \ce{CH_4} in this disk. The tentative detection is further highlighted in Section \ref{sec:CH4}. The molecular species are highlighted in Figure \ref{fig:CYTau-ProfFit}.  \\
\indent \textit{DN~Tau}: DN~Tau has previously been classified, just as CX~Tau, as a \ce{CO_2}-rich source \citep{PontoppidanEA10}. Aside from the strong emission of \ce{CO_2}, we report the detection of its isotopologue \ce{^{13}CO_2} and the potential detection of \ce{CO^{18}O}. Other molecular species, such as \ce{HCN} and \ce{C_2H_2} are not confidently detected, while \ce{OH} is. At the shortest wavelengths, \ce{CO} is only tentatively detected, while emission from the ro-vibrational \ce{H_2O} transitions is clearly detected. As for the other sources, both the \ce{H_2} S(1) and S(2) transitions are also detected. Figure \ref{fig:DNTau-ProfFit} highlights the detected molecular species. \\
\indent \textit{FT~Tau}: Aside from the strong pure rotational \ce{H_2O} emission, the spectrum of FT~Tau also contains emission from the ro-vibrational lines and from \ce{CO}. At the longer wavelengths, we clearly detect \ce{OH}, \ce{CO_2}, \ce{HCN}, and \ce{C_2H_2}. The \ce{H_2} S(1) and S(2) transitions are also detected. The emission of these species is also highlighted in Figure \ref{fig:FTTau-ProfFit}. \\
\indent \textit{RNO~90}: Besides the rotational \ce{H_2O} transitions, the spectrum of RNO~90 also contains emission from \ce{CO}, the ro-vibrational \ce{H_2O} transitions, \ce{OH}, \ce{CO_2}, \ce{HCN}, and \ce{C_2H_2}. Both the S(1) and S(2) transitions of \ce{H_2} are also detected. The molecular emission is highlighted in Figure \ref{fig:RNO90-ProfFit}. \\
\indent \textit{XX~Cha}: Although the spectrum of XX~Cha is very similar to that of FT~Tau, the emission from both \ce{CO_2} and \ce{C_2H_2} is stronger. Emission from \ce{CO}, the ro-vibrational \ce{H_2O} lines, \ce{OH}, and \ce{HCN} are also clearly detected. As for all the other sources, both the \ce{H_2} S(1) and S(2) transitions are also detected. Figure \ref{fig:XXCha-ProfFit} highlights the molecular emission of the other species in the spectrum of XX~Cha. 

\subsection{Tentative detection of \ce{CH_4} in CY~Tau} \label{sec:CH4}
\begin{figure}[ht!]
    \centering
    \includegraphics[width=0.9\textwidth]{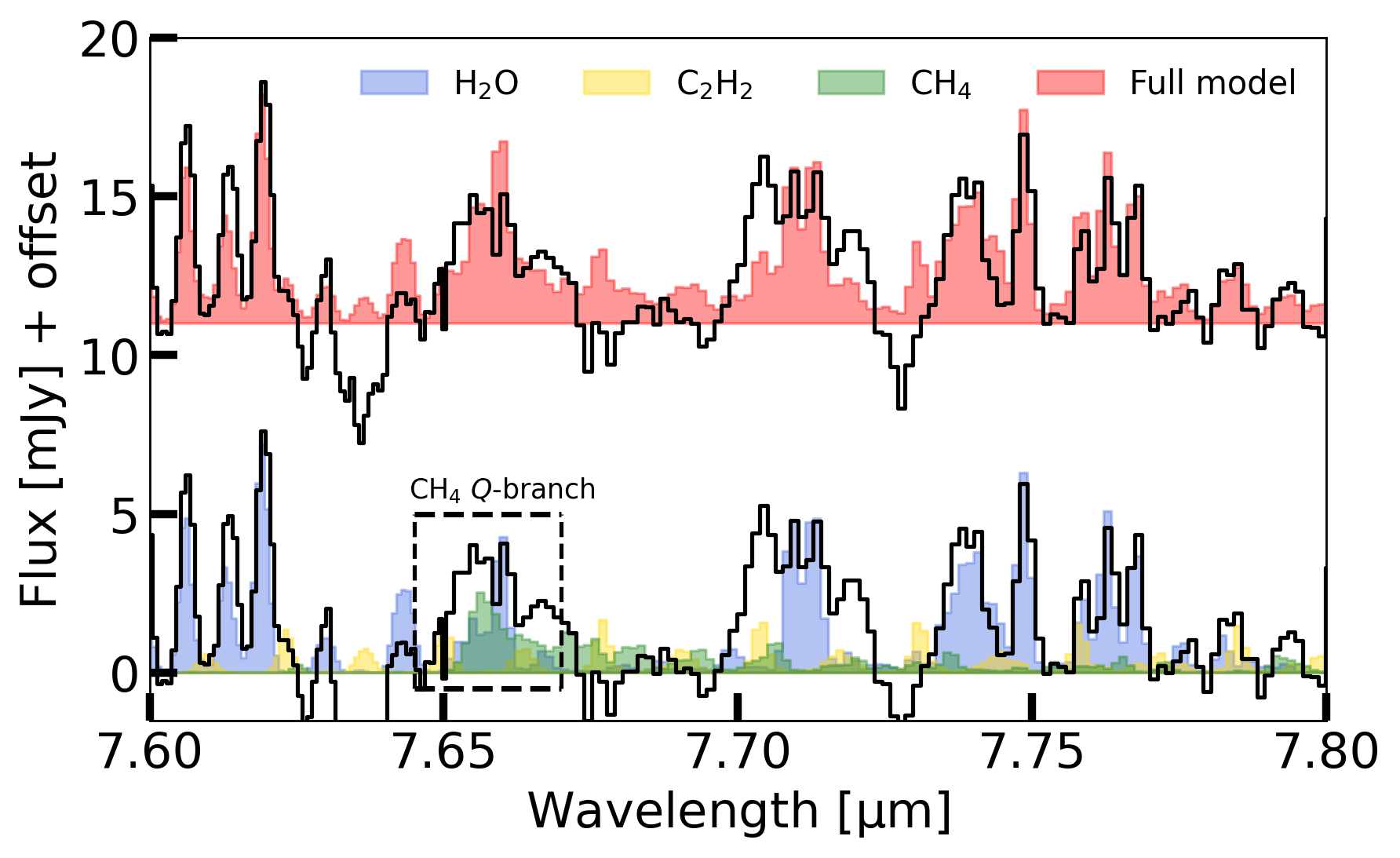}
    \caption{Zoom-in on the 7.60-8.00 $\mathrm{\mu}$m region of CY~Tau, showing the potential detection of \ce{CH_4} (green). Also shown are approximate contributions from \ce{H_2O} (blue) and \ce{C_2H_2} (yellow) slab models. The full model is shown in red with a small offset applied to the flux.}
    \label{fig:CH4-Angel}
\end{figure}
We report a tentative detection of \ce{CH_4} in the spectrum of CY~Tau, whose $Q$-branch is located at $\sim$7.66 $\mathrm{\mu}$m. In Figure \ref{fig:CH4-Angel}, we display the 7.60-8.00 $\mathrm{\mu}$m wavelength region, containing molecular emission from ro-vibrational \ce{H_2O} transitions and potential weak features from \ce{C_2H_2} and the potential detection of \ce{CH_4}. For the slab models (\ce{H_2O}, \ce{C_2H_2}, and \ce{CH_4}) we have used excitation temperatures of 975 K, 500 K, and 500 K, column densities ($\log_{10}(N)$ with $N$ in cm$^{-2}$) of 18.4, 17.0, and 16.8, and emitting radii of 0.03 au, 0.10 au, and 0.10 au, respectively. The parameters for \ce{H_2O} and \ce{C_2H_2} were obtained using a $\chi_\mathrm{red}^2$-approach similar to that described in Section \ref{sec:AnalH2O} but using grids instead of an MCMC exploration (see also, for example, \citealt{GrantEA23}). The model for \ce{CH_4} has been obtained by visually fitting the residuals. These slab models thus suggest a column density ratio of $N_\mathrm{\ce{C_2H_2}}/N_\mathrm{\ce{CH_4}}\gtrsim$1.5, under the assumption that the species are co-existing. The potential presence of \ce{CH_4}, in addition to the detection of \ce{^{13}CCH_2}, suggests that CY~Tau may be the most carbon-rich disk in our sample. This disk may be similar to the much larger disk of DoAr~33, where \citet{ColmenaresEA24} reported, alongside other hydrocarbons, a tentative detection of \ce{CH_4}. They compared their spectra to thermochemical models and argued for an overall carbon-to-oxygen ratio (\ce{C}/\ce{O}) larger than unity.

\section{Integrated line fluxes}
\begin{table}[ht!]
    \centering
    \caption{Integrated fluxes for the lines tracing the cold, intermediate, and hot \ce{H_2O} reservoirs.}
    \begin{tabular}{c c c c c | c}
        \hline
        Source & $F_\mathrm{1448K}$ & $F_\mathrm{1615K}$ & $F_\mathrm{3646K}$ & $F_\mathrm{6052K}$ & $F_\mathrm{1500K}^\alpha$ \\
        \hline
        BP~Tau & 10.67$\pm$0.07 & 10.43$\pm$0.04 & 6.68$\pm$0.05 & 3.15$\pm$0.05 & 21.09$\pm$0.08 \\
        CX~Tau & 0.99$\pm$0.02 & 0.93$\pm$0.02 & 0.49$\pm$0.03 & 0.09$\pm$0.03 & 1.91$\pm$0.03 \\
        CY~Tau & 1.17$\pm$0.01 & 1.22$\pm$0.01 & 1.02$\pm$0.01 & 0.60$\pm$0.01 & 2.40$\pm$0.01 \\
        DN~Tau & 1.29$\pm$0.03 & 1.49$\pm$0.05 & 0.57$\pm$0.02 & 0.13$\pm$0.02 & 2.78$\pm$0.06 \\
        DR~Tau & 36.38$\pm$0.20 & 33.89$\pm$0.16 & 24.94$\pm$3.17 & 13.49$\pm$0.26 & 70.27$\pm$0.26 \\
        FT~Tau & 8.60$\pm$0.03 & 6.99$\pm$0.03 & 2.81$\pm$0.03 & 2.13$\pm$0.03 & 15.59$\pm$0.05 \\
        RNO~90 & 66.03$\pm$0.26 & 66.65$\pm$0.22 & 40.82$\pm$0.19 & 24.72$\pm$0.27 & 132.68$\pm$0.34 \\
        XX~Cha & 4.23$\pm$0.01 & 3.88$\pm$0.02 & 1.54$\pm$0.01 & 1.09$\pm$0.01 & 8.11$\pm$0.02 \\
        \hline
    \end{tabular}
    \label{tab:LineFluxes}
    \tablefoot{The fluxes are given in $\times10^{-15}$ erg~s$^{-1}$~cm$^{-2}$. \\
    $^\alpha$: The $F_\mathrm{1500K}$ flux is the summation of the fluxes of the 1448 K and 1615 K lines, following \citet{BanzattiEA24}.}
\end{table}

\clearpage
\section{Profile comparison} \label{sec:ProfComp}
\begin{figure}[ht!]
    \centering
    \includegraphics[width=0.9\linewidth]{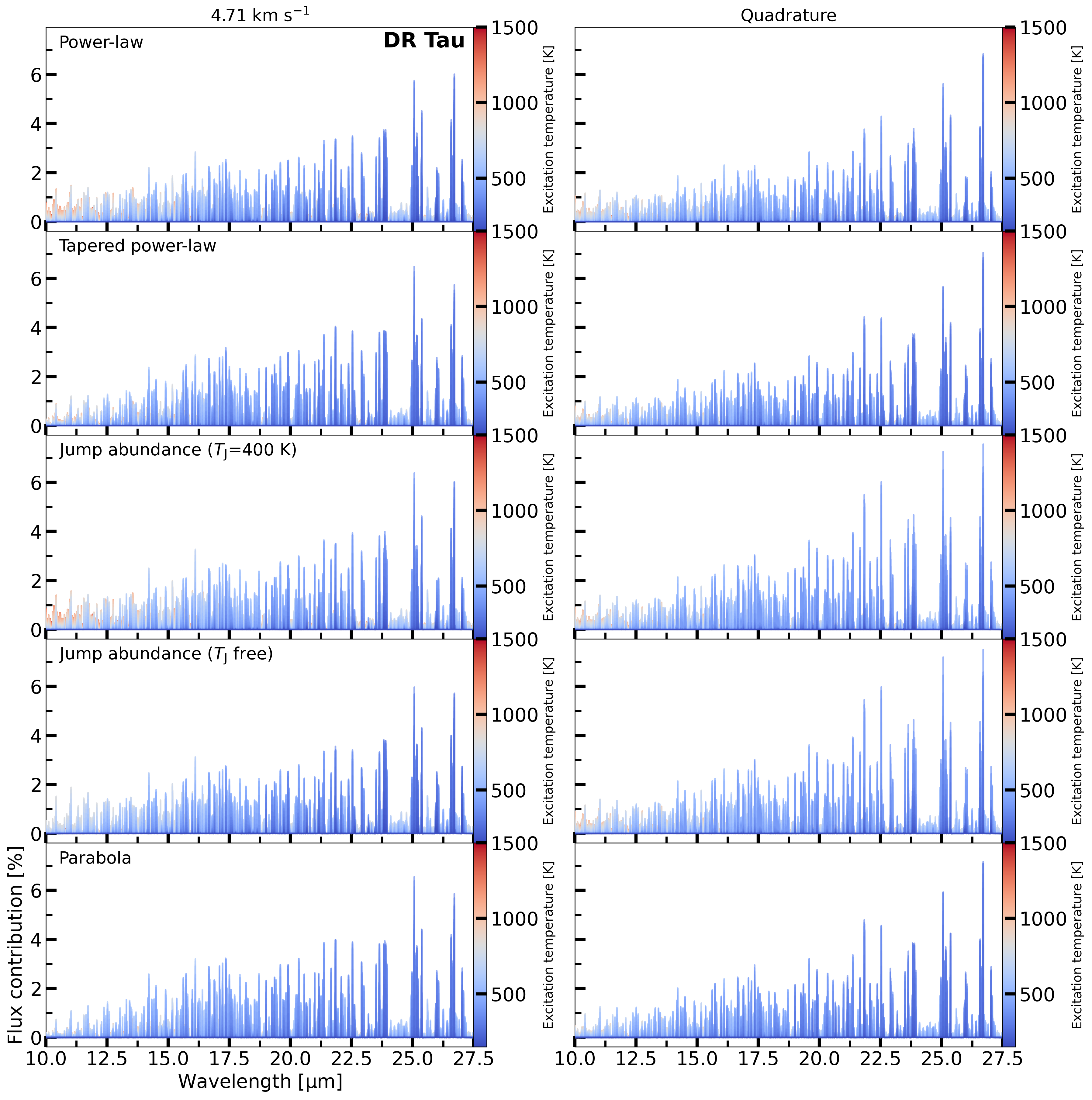}
    \caption{The contributions from all parametric fits for DR~Tau, representing the Type N disks.}
    \label{fig:Contr-DRTau}
\end{figure}

\begin{figure}[ht!]
    \centering
    \includegraphics[width=0.9\linewidth]{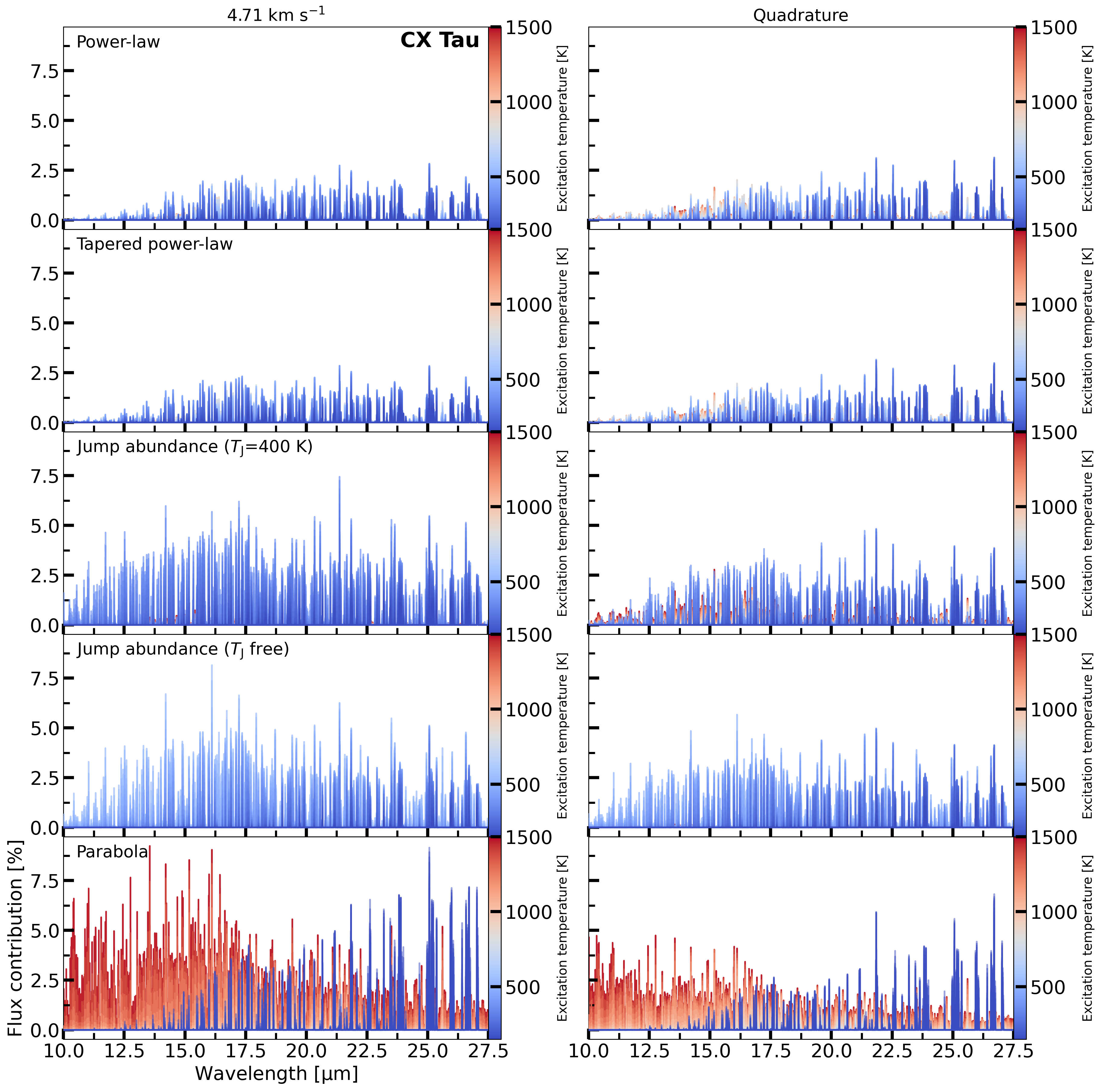}
    \caption{Similar as Figure \ref{fig:Contr-DRTau}, but for CX~Tau, representing the Type P disks.}
    \label{fig:Contr-CXTau}
\end{figure}

\begin{figure}[ht!]
    \centering
    \includegraphics[width=0.9\linewidth]{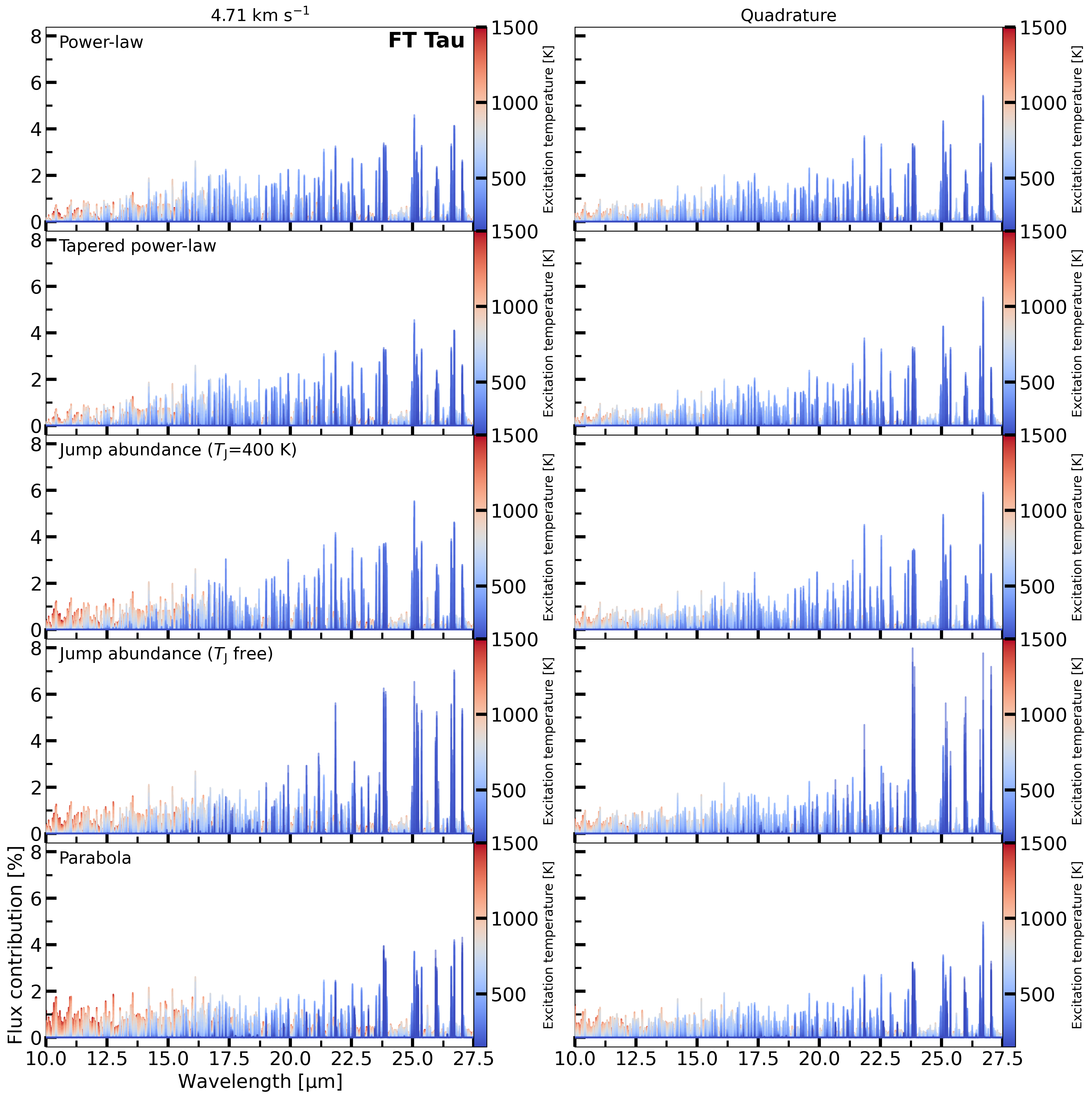}
    \caption{Similar as Figure \ref{fig:Contr-DRTau}, but for FT~Tau, representing the Type E disks.}
    \label{fig:Contr-FTTau}
\end{figure}

In Figures \ref{fig:Contr-DRTau}, \ref{fig:Contr-CXTau}, and \ref{fig:Contr-FTTau} we show the flux contributions of the 50 slab models for the profiles fitted for DR~Tau, CX~Tau, and FT~Tau, respectively. These disks are each part of one of the different types and comparing their contribution plots provides information on how well the different profiles fit for their types. \\
\indent \textit{DR~Tau (Type N)}: The Type N disks (BP~Tau, CY~Tau, DR~Tau, and RNO~90) are overall best described by the exponentially tapered power laws, the jump abundances with the jump occurring at high temperatures, and upward parabola (see also Table \ref{tab:ResultsSum}). As can be seen in Figure \ref{fig:Contr-DRTau}, those profiles all yield very similar contributions. The main differences between those profiles and the ones with the simple power law or the jump abundance with the jump fixed at $T_\mathrm{jump}$=400 K is the flux at the shortest wavelengths ($<$15.0 $\mathrm{\mu}$m), best probed by the hottest component, where the power law and the fixed temperature jump abundance have stronger contributions of the hotter components (most notable in the left panels of Figure \ref{fig:Contr-DRTau}. These stronger contributions results in overfitting the observed flux in those inner regions, whereas the preferred profiles have a decrease in column density in the innermost and, therefore, do not overfit these inner regions. \\
\indent \textit{CX~Tau (Type P)}: The Type P disks (CX~Tau and DN~Tau) mainly prefer the downward parabola and jump abundance profiles (see also Table \ref{tab:ResultsSum}). As can be seen in Figure \ref{fig:Contr-CXTau}, those profiles have the strongest contributions from either the hot or intermediate and cold components, while the (tapered) power law profiles are not able to capture the strength of all three components (hot, intermediate, and cold). The preference for these profiles shows that the P Type disks do have contributions from at least two of the components.  \\
\indent \textit{FT~Tau (Type E)}: FT~Tau and XX~Cha, the Type E disks, are best described by a jump abundance at low temperatures or a smooth power law (see also Table \ref{tab:ResultsSum}). As can be seen in Figure \ref{fig:Contr-FTTau} for FT~Tau, the jump abundance with a jump at $T_\mathrm{jump}\lesssim$250 K is really able to capture the strength of the cold component. The power law is not able to capture this cold component as strongly, but the contributions are clearly still there. The parabola, while being able to yield a similar contribution to the cold component as the power law, has a much stronger contribution at the shortest wavelengths ($<$15.0 $\mathrm{\mu}$m) following its upward turn. This stronger contribution actually overfits the observed flux at these wavelengths, disfavouring the parabola as the best fit.


\end{appendix}

\end{document}